\documentclass[prl,aps,twocolumn,superscriptaddress,floatfix,noshowpacs,10pt]{revtex4-2}

\usepackage[utf8]{inputenc}
\usepackage{microtype}
\usepackage{amsmath,amssymb}
\usepackage{graphicx}
\usepackage{ulem}
\usepackage[colorlinks=true]{hyperref}

\usepackage{bm}
\renewcommand{\vec}{\bm}
\renewcommand{\Re}{\operatorname{Re}}
\renewcommand{\Im}{\operatorname{Im}}

\newcommand{\dif}{\mathrm{d}}
\newcommand{\mi}{\mathrm{i}}
\newcommand{\me}{\mathrm{e}}

\begin{document}

\title{Time Translation Symmetry Breaking in an Isolated Spin-Orbit-Coupled Fluid of Light}

\author{Giovanni I. Martone}
\email{giovanni\textunderscore italo.martone@lkb.upmc.fr}
\affiliation{Laboratoire Kastler Brossel,
Sorbonne Universit\'{e}, CNRS, ENS-PSL Research University, 
Coll\`{e}ge de France; 4 Place Jussieu, 75005 Paris, France}
\affiliation{CNR NANOTEC, Institute of Nanotechnology,
Via Monteroni, 73100 Lecce, Italy}
\affiliation{INFN, Sezione di Lecce, 73100 Lecce, Italy}

\author{Nicolas Cherroret}
\email{nicolas.cherroret@lkb.upmc.fr}
\affiliation{Laboratoire Kastler Brossel,
Sorbonne Universit\'{e}, CNRS, ENS-PSL Research University, 
Coll\`{e}ge de France; 4 Place Jussieu, 75005 Paris, France}

\begin{abstract}
We study the interplay between intrinsic spin-orbit coupling and nonlinear photon-photon interactions in a nonparaxial, elliptically polarized fluid
of light propagating in a bulk Kerr medium. We find that in situations where the nonlinear interactions induce birefringence, i.e., a
polarization-dependent nonlinear refractive index, their interplay with spin-orbit coupling results in an interference between the two polarization
components of the fluid traveling at different wave vectors, which entails the breaking of translation symmetry along the propagation direction. This
phenomenon leads to a Floquet band structure in the Bogoliubov spectrum of the fluid, and to characteristic oscillations of its intensity correlations.
We characterize these oscillations in detail and point out their exponential growth at large propagation distances, revealing the presence of parametric
resonances.
\end{abstract}

\maketitle

Spin-orbit coupling (SOC) in materials arises due to the interaction between the electron spin and its momentum, and lies at the heart of various
phenomena and concepts such as spin Hall effects~\cite{Kato2004,Wunderlich2005}, topological insulators~\cite{Hasan2010}, and Majorana
fermions~\cite{Sau2011}. In the context of quantum fluids, progress in the engineering of synthetic gauge fields has paved the way for
intriguing phenomena resulting from the interplay between SOC and particle interactions in ultracold atoms~\cite{Goldman2014,Zhai2015}. In the
ground states of Bose gases, e.g., this interplay yields stripe-superfluid or lattice phases~\cite{Wang2010,Ho2011,Wu2011,Sinha2011,Li2012,
Li2016}. In degenerate Fermi gases, on the other hand, SOC can significantly impact the celebrated BEC-BCS crossover~\cite{Vyasanakere2011} or lead
to topological superfluids~\cite{Qu2013,Zhang2013}. Beyond matter waves, SOC also exists in photonic systems~\cite{Shelykh2010,Carusotto2013}.
This has been demonstrated, in particular, for exciton-polaritons in microcavities~\cite{Kavokin2005,Leyder2007,Tercas2014,Gianfrate2020,
Polimeno2021a,Polimeno2021b} or for photons tunneling in properly designed microstructures~\cite{Sala2015}. In those systems, effective photon-photon
interactions also emerge due to the interaction between the underlying excitons, and their interplay with SOC has been investigated in
numerous works~\cite{Shelykh2006,Flayac2010,Hivet2012,Flayac2013,Bleu2016}.

An alternative optical platform where photon interactions can be realized are fluids of light in the propagating geometry~\cite{Carusotto2014}.
Here the propagation of light through a nonlinear medium mimicks, in the paraxial limit, the temporal evolution of a two-dimensional (2D)
quantum fluid, the propagation axis playing the role of an effective time and the nonlinearity mediating the photon interactions. This analogy has
been beautifully illustrated with measurements of the Bogoliubov dispersion~\cite{Vocke2015,Fontaine2018}, the dynamical formation of optical
condensates~\cite{Sun2012,Santic2018}, the spontaneous nucleation of vortices in a photonic lattice~\cite{Situ2020}, or the temporal dynamics of
correlation functions following a quench~\cite{Steinhauer2022,Abuzarli2022}. Owing to the absence of cavity or underlying microstructure, fluids of
light in the propagating geometry do not apparently seem to constitute a natural platform for achieving SOC. Nevertheless, recently a spin-orbit
mechanism has been demonstrated in this system~\cite{Martone2021}, based on the fundamental coupling between the polarization and the trajectory
of optical fields subjected to a refractive-index gradient. Unlike linear setups, where the gradient is provided by the medium
inhomogeneity~\cite{Onoda2004,Hosten2008,Bliokh2015,Ling2017,Jisha2017,Cherroret2018,BardonBrun2019,Zhang2020}, in nonlinear media SOC emerges in the
presence of fairly strong spatial variations of the optical field itself, requiring it to deviate from its paraxial propagation regime. This induces a
nonlinear index gradient, which couples to the optical spin via the polarization-trajectory coupling term of the wave equation~\cite{Martone2021}.

\begin{figure}
\includegraphics[scale=0.9]{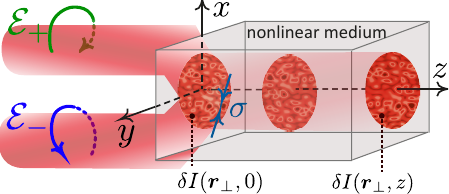}
\caption{We study the propagation of an optical field with two
circularly polarized components $\mathcal{E}_\pm$ in a
nonlinear medium. Both $\mathcal{E}_+$ and $\mathcal{E}_-$ display weak
spatial intensity fluctuations $\delta I(\vec{r}_\perp,z=0)$ (modeled by a
speckle with correlation length $\sigma$), which couple to the optical spin
via the nonlinear refractive index. Breaking of translation invariance
along $z$ is observed in the input-output correlation
$\langle\delta I(\vec{r}_\perp,0)\delta I(\vec{r}_\perp,z)\rangle$.}
\label{fig:setup}
\end{figure}

In this Letter, we show that elliptically polarized fluids of light propagating in media displaying nonlinear birefringence exhibit a breaking of
translation symmetry along the optical axis direction (i.e., the effective-time axis) due to SOC. In stark contrast with the case of linearly
polarized fluids considered in Ref.~\cite{Martone2021}, this leads to the emergence of a Floquet band structure in the excitation spectrum,
analogously to what is observed in driven systems~\cite{Holthaus2016}, but here in a purely \textit{isolated} optical fluid. The breaking of
translation symmetry also gives rise to peculiar oscillations in several physical quantities, in particular the intensity-correlation function of
the fluid of light. By characterizing these oscillations in detail, we further point out their exponential growth at large propagation distances.
This showcases the existence of parametric resonances, an original manifestation of the interplay between nonlinearity and SOC in fluid systems.

Our setup consists of a bulk nonlinear medium infinitely extended along the $x$ and $y$ axes and $z > 0$ (see Fig.~\ref{fig:setup}).
A monochromatic field $\vec{E}(\vec{r},t) = \Re \left[ \vec{\mathcal{E}}(\vec{r}) \me^{- \mi \omega_0 t} \right]$ propagates inside the material at
frequency $\omega_0$. The components of the complex amplitude $\vec{\mathcal{E}}$ obey the nonlinear Helmholtz equation
\begin{equation}
\nabla^2 \vec{\mathcal{E}} - \nabla \left( \nabla \cdot \vec{\mathcal{E}} \right)
+ \frac{\omega_0^2}{c^2} \, \left[n_0^2 + 2 n_0 \Delta n(\vec{\mathcal{E}}) \right] \vec{\mathcal{E}} = 0 \, ,
\label{eq:helm_eq}
\end{equation}
with $c$ the vacuum speed of light and $n_0$ the linear refractive index. The nonlinear refractive index $\left[ \Delta n(\vec{\mathcal{E}})
\right]_{ij} = \left(n_{2,d} + n_{2,s}\right) |\vec{\mathcal{E}}|^2 \delta_{ij} - n_{2,s} \mathcal{E}_i^* \mathcal{E}_j$ ($i,j=x,y,z$) is a tensor
featuring two independent Kerr indices $n_{2,d}$ and $n_{2,s}$. Equation~\eqref{eq:helm_eq} corresponds to the Euler-Lagrange equation for the action
functional $\mathcal{S} = \int \dif^3 \vec{r} \, \mathcal{L}$, where~\cite{Martone2021}
\begin{equation}
\begin{split}
\mathcal{L} = {}&{} - \frac{1}{2 \beta_0} \left(\nabla_i \mathcal{E}_j^* \nabla_i \mathcal{E}_j - \nabla_i \mathcal{E}_j^* \nabla_j \mathcal{E}_i
- \beta_0^2 \mathcal{E}_i^* \mathcal{E}_i\right) \\
&{} - \frac{1}{2} \left[ g_d \delta_{ij} \delta_{i'j'} + g_s \left( S_k \right)_{ij} \left( S_k \right)_{i'j'} \right]
\mathcal{E}_i^* \mathcal{E}_j \mathcal{E}_{i'}^* \mathcal{E}_{j'} \, .
\end{split}
\label{eq:helm_lag}
\end{equation}
Here summation over repeated indices is implied, $\beta_0 = n_0 \omega_0 / c$ is the propagation constant, and $(S_k)_{ij} = - \mi \varepsilon_{ijk}$
denotes the $k^\text{th}$ spin-$1$ matrix. The simultaneous presence of a spin-independent ($g_d = - n_{2,d} \omega_0 / c$) and a spin-dependent
($g_s = - n_{2,s} \omega_0 / c$) nonlinear coupling, both assumed positive, is typical of isotropic systems described by multicomponent fields, where
the pairing in channels of different total spin can occur at different strengths. Besides optical nonlinear media~\cite{Landau_Lifshitz_08_book,
Agrawal_book}, this behavior is observed in atomic spinor Bose-Einstein condensates~\cite{StamperKurn2013review} and in microcavity
exciton-polaritons~\cite{Shelykh2004,Shelykh2010,Johne2010,Stepanov2019}, where the spin-dependent term is sometimes called self-induced Zeeman splitting.
Note, however, that unlike polariton condensates~\cite{Shelykh2006,Flayac2010,Hivet2012,Flayac2013,Bleu2016}, the fluid of light described by
Eq.~\eqref{eq:helm_eq} is fully three-dimensional. In our setup, the mechanism of SOC of light originates from the term $\nabla \left( \nabla \cdot
\vec{\mathcal{E}} \right) \sim \nabla \left(\nabla\ln \Delta n\cdot\vec{\mathcal{E}}\right)$ in Eq.~\eqref{eq:helm_eq}, which couples the fluid
polarization to its trajectory (via the nonlinear index gradient). In the corresponding Lagrangian formalism, Eq.~\eqref{eq:helm_lag},
the SOC effects are encoded in the term $\propto \nabla_i \mathcal{E}_j^* \nabla_j \mathcal{E}_i$. While naturally present in Maxwell equations, the latter
is discarded within the usual paraxial approximation~\cite{Carusotto2014}. In the following, we do \textit{not} perform this approximation but work
out the full Lagrangian [Eq.~\eqref{eq:helm_lag}].

Our aim is to determine the field amplitude inside the medium, given its transverse profile $\vec{\mathcal{E}}[\vec{r}_\perp=(x,y),z=0]$ at the
air-medium interface. This can be regarded as the evolution problem of a 2D system with respect to the effective time $z$ \cite{Carusotto2014}. In the
following, we assume that $\vec{\mathcal{E}}$ is the sum of a large homogeneous background and a small fluctuation, and treat the latter using Bogoliubov-Popov
theory~\cite{Popov1972,Popov_book,Mora2003,Petrov2004}. For that purpose, we write $\vec{\mathcal{E}}=(\mathcal{E}_+,\mathcal{E}_z,\mathcal{E}_-)^T$ and employ
the density-phase decomposition $\mathcal{E}_+ = \sqrt{I} \cos (\vartheta/2) \, \me^{\mi (\Theta + \chi / 2)}$,  $\mathcal{E}_- = \sqrt{I} \sin (\vartheta/2)
\, \me^{\mi (\Theta - \chi / 2)}$ of the field circular components $\mathcal{E}_\pm = \mp (\mathcal{E}_x \mp \mi \mathcal{E}_y) / \sqrt{2}$. Here $I$ and
$\vartheta$ quantify the total optical intensity of the transverse components and their relative weight, respectively, while $\Theta$ ($\chi$) is their total
(relative) phase. We then split the field into a background and a fluctuating contribution, writing $I = I_0 + \delta I$, $\vartheta = \vartheta_0
+ \delta \vartheta$, and $\chi = \Delta k \:\! z+\delta \chi$, where $\Delta k=k_+-k_-$. The wave numbers $k_\pm$ of the two polarization components are imposed
by Eq.~\eqref{eq:helm_eq}:
\begin{equation}
k_\pm= \sqrt{\beta_0^2 - 2 \beta_0 (g_d \pm g_s \cos\vartheta_0) I_0} \, .
\label{eq:refr_ind}
\end{equation}
Notice that $\Delta k = k_+-k_- \neq 0$ as soon as $g_s \neq 0$ and $\cos\vartheta_0 \neq 0$, i.e., when the background field is elliptically or
circularly polarized. This defines the phenomenon of nonlinear circular birefringence, which will play a crucial role in the following.

Next, we insert the fluctuation variables into the Lagrangian [Eq.~\eqref{eq:helm_lag}] and determine the quadratic correction $\mathcal{S}^{(2)}$
to the background action. This is achieved by redefining $\mathcal{E}_z \to \me^{\mi \Theta} \mathcal{E}_z$ and $\Theta\to \Theta+(k_++k_-)z/2$, and
expanding Eq.~\eqref{eq:helm_lag} with respect to $\delta I$, $\delta \vartheta$, and $\delta \chi$. Note that the fluctuations of $\Theta$ are, in
contrast, possibly large in two dimensions~\cite{Mora2003}, but its derivatives remain small, and so does $\mathcal{E}_z$. In this procedure,
$\mathcal{S}^{(2)}$ turns out to be independent of the $z$ derivatives of $\mathcal{E}_z$, so that one can use the Euler-Lagrange equation
$\delta \mathcal{S}^{(2)} / \delta \mathcal{E}_z^* = 0$ to eliminate $\mathcal{E}_z$~\cite{Supplemental}. The quadratic action $\mathcal{S}^{(2)}
= \int \dif z \int {\dif^2 \vec{q}_\perp}/{(2\pi)^2} \tilde{\mathcal{L}}^{(2)}$ can be written in terms of a single column vector
$X = (\delta\tilde{I}/2I_0, \delta\tilde{\vartheta}/2, \tilde{\Theta}, \delta\tilde{\chi}/2)^T$ for the Fourier variables with respect
to $\vec{r}_\perp$, e.g., $\delta \tilde{I}(\vec{q}_\perp,z) = \int \dif^2 \vec{r}_\perp \, \delta I(\vec{r}_\perp,z) \, \me^{- \mi \vec{q}_\perp \cdot
\vec{r}_\perp}$, with $\vec{q}_\perp = \left(q_\perp\cos\varphi_q,q_\perp\sin\varphi_q\right)$ the transverse momentum. We find that
\begin{equation}
\tilde{\mathcal{L}}^{(2)} = \dot{X}^\dagger \Lambda_2 \dot{X} + \dot{X}^\dagger \Lambda_1 X
+ X^\dagger \Lambda_1^T \dot{X} - X^\dagger \Lambda_0 X \, ,
\label{eq:lagr_2} 
\end{equation}
where $\dot{X}\equiv\partial_zX$. The $4 \times 4$ matrices $\Lambda_{0,1,2}$ are real $\pi$-periodic functions of the angular variable $\varphi(z)
= \varphi_q + \Delta k \:\! z / 2$~\cite{Supplemental}, which encodes a breaking of translation invariance along the effective time axis $z$, the central
result of the Letter. This dependence stems from the SOC term in Eq.~\eqref{eq:helm_lag}, and is completely absent in the paraxial framework. In our
description, the paraxial approximation corresponds to taking $\dot{X}/\beta_0$, $(q_\perp/\beta_0)^2$, and $g_{d,s}I_0 / \beta_0$ small; the resulting
expansion of $\tilde{\mathcal{L}}^{(2)}$ up to first order becomes $\varphi(z)$ independent and formally identical to the Bogoliubov Lagrangian of symmetric
binary mixtures of atomic condensates~\cite{Pethick_Smith_book,Pitaevskii_Stringari_book}.

In the same spirit as in Ref.~\cite{Martone2021}, we define a Hamiltonian $\tilde{\mathcal{H}}^{(2)} = \Pi^T \dot{X} + \dot{X}^\dagger \Pi^* -
\tilde{\mathcal{L}}^{(2)}$ depending on $X$ and the conjugate momenta vector $\Pi = \partial \tilde{\mathcal{L}}^{(2)} / \partial \dot{X}^T$.
The effective-time evolution of $X$ and $\Pi$ is governed by the Hamilton equations $\dot{X} = \partial \tilde{\mathcal{H}}^{(2)} / \partial \Pi^T$
and $\dot{\Pi} = - \partial \tilde{\mathcal{H}}^{(2)} / \partial X^T$, yielding the eight coupled equations
\begin{equation}
\begin{bmatrix}
\dot{X} \\
\dot{\Pi}^*
\end{bmatrix}	
=
\begin{bmatrix}
- \Lambda_2^{-1} \Lambda_1 & \Lambda_2^{-1} \\
- (\Lambda_1^T \Lambda_2^{-1} \Lambda_1 + \Lambda_0) & (\Lambda_2^{-1} \Lambda_1)^T
\end{bmatrix}
\begin{bmatrix}
X \\
\Pi^*
\end{bmatrix} \, .
\label{eq:ham_eq}
\end{equation}
In Ref.~\cite{Martone2021}, Eq.~\eqref{eq:ham_eq} was solved in the $\Delta k = 0$ case (linearly polarized background field), where the matrix of
coefficients is constant. Here on the contrary, we assume that the birefringence condition $\Delta k \neq 0$ is fulfilled. Hence, the coefficients of
Eq.~\eqref{eq:ham_eq} oscillate in $z$ with period $2 \pi / \Delta k$. We stress that although this behavior is typical of Bogoliubov equations for
periodically driven systems~\cite{Tozzo2005,Creffield2009,Bukov2015,Lellouch2017,Boulier2019} (see also Ref.~\cite{Eckardt2017}), here it occurs in a
purely \textit{isolated} system due to the breaking of time-translation symmetry by the background. The underlying mechanism is the interplay between
nonlinear birefringence and SOC. The latter is responsible for the presence, in Eq.~\eqref{eq:helm_lag}, of interference terms between the two
polarization components propagating at relative wave vector $\Delta k$. Note that this phenomenon is absent in linear birefringent media, where the
fluctuation is always independent of the background field.

According to Floquet's theorem~\cite{Floquet1883,Chicone_book}, the general solution of Eq.~\eqref{eq:ham_eq} has the form
\begin{equation}
\begin{bmatrix}
X(\vec{q}_\perp,z) \\
\Pi^*(\vec{q}_\perp,z)
\end{bmatrix}
= \sum_\ell C_\ell(\vec{q}_\perp)
\begin{bmatrix}
X_{0,\ell}(q_\perp,\varphi) \\
\Pi_{0,\ell}^*(q_\perp,\varphi)
\end{bmatrix}
\me^{- \mi \Omega_\ell(q_\perp) z}.
\label{eq:speckle_gen_sol}
\end{equation}
Here the sum runs over eight independent solutions, labeled by $\ell$ and appearing with weight $C_\ell$. These Floquet solutions are characterized by
their eigenfunctions $X_{0,\ell}$ and $\Pi_{0,\ell}^*$ and the corresponding quasifrequencies, $\Omega_\ell$. Note that as is customary for Bogoliubov
equations~\cite{Castin_review}, for each solution with quasifrequency $\Omega_\ell$ there exists another one with quasifrequency $- \Omega_\ell^*$
associated with the same physical oscillation, hence a total of four Bogoliubov modes. In the paraxial regime, in contrast, one has only a density ($d$)
and a spin ($s$) mode, characterized by in- and out-of-phase intensity oscillations of the two polarization components, respectively, as discussed in
the Supplemental Material~\cite{Supplemental}. 

\begin{figure}
\includegraphics[scale=0.15]{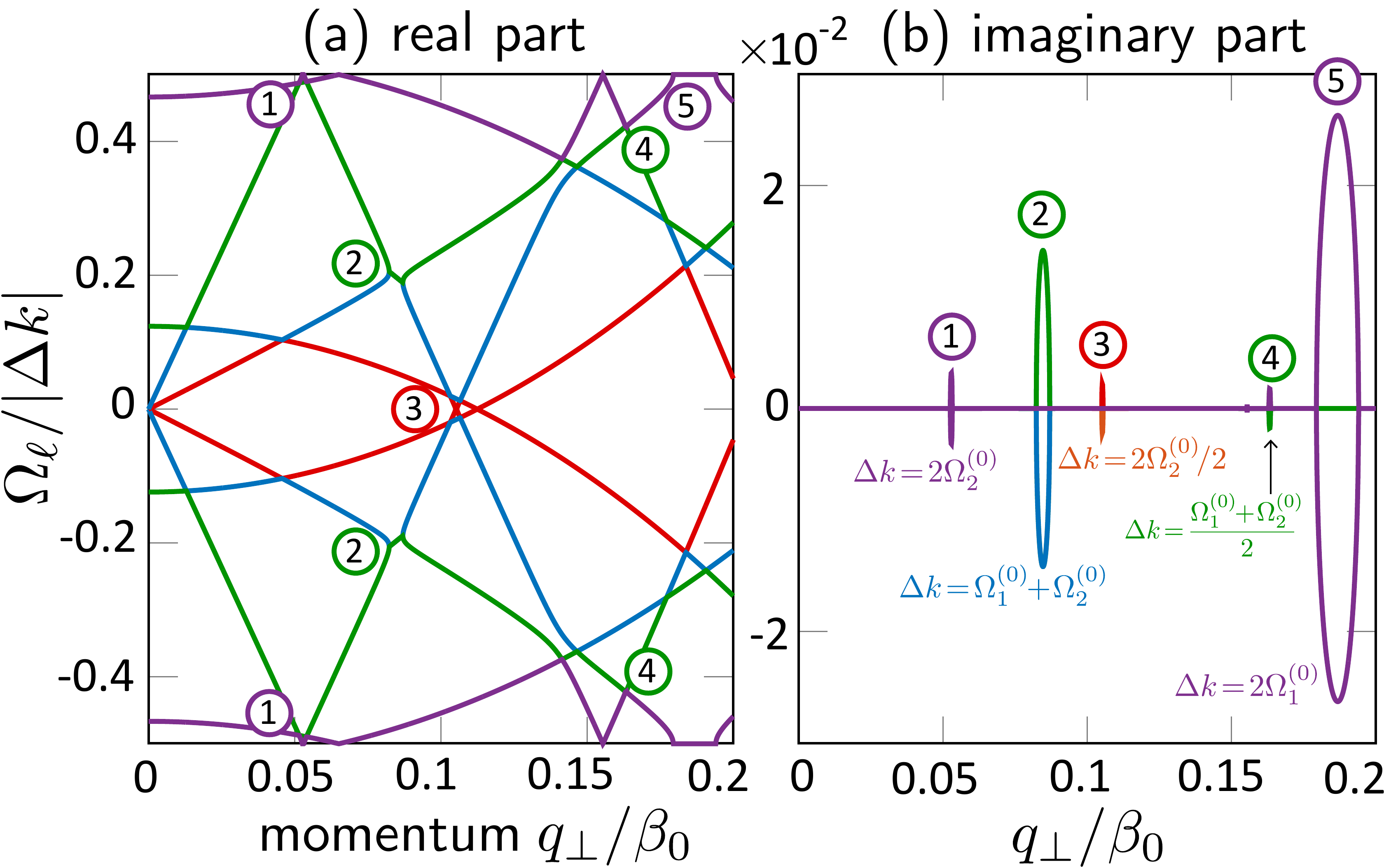}
\caption{(a) Real and (b) imaginary parts of the quasifrequency
spectrum of an elliptically polarized fluid of light as functions of
the transverse momentum. Each parametric resonance is assigned a number,
which is used to identify the corresponding regions in the two plots.
In (b) the corresponding resonance conditions are also provided.
Here we have chosen the background polarization $\vartheta_0 = \pi / 4$ and
the nonlinear couplings $g_d I_0 / \beta_0 = 0.2$,
$g_s I_0 / \beta_0 = 0.05$.}
\label{fig:omega_ell}
\end{figure}

We first plot the real part of the quasifrequency spectrum in Fig.~\ref{fig:omega_ell}(a). Because the $\Omega_\ell$'s are defined modulo $\Delta k$,
it is sufficient to take their real part in the first Brillouin zone, $-|\Delta k|/2 < \Re \Omega_\ell \leq |\Delta k|/2$ \cite{Supplemental}.
In the $q_\perp \to 0$ limit, the spectrum in panel (a) displays the usual four phononic bands $\pm \Omega_{d,s}(q_\perp) \simeq \pm c_{d,s} q_\perp$,
with two sound velocities $c_{d,s}$. Those correspond to the standard Bogoliubov modes in the paraxial regime, where
$c_{d(s)}^2 = (g_d + g_s \pm \sqrt{g_d^2 + g_s^2 + 2 g_d g_s \cos 2\vartheta_0}) I_0 / 2\beta_0$. On the contrary, the other four bands tend to a
finite value and describe light propagating backward along $z$. At increasing $q_\perp$ the various bands cross one another at several points, giving
rise to an involved structure. In addition, some quasifrequencies---labeled by numbers in Fig.~\ref{fig:omega_ell}(a)---develop a finite imaginary
part at certain values of $q_\perp$, see Fig.~\ref{fig:omega_ell}(b) (notice that complex quasifrequencies always occur in complex conjugate pairs).
This important result reveals the presence of parametric resonances, which will be discussed in more details below.

A central feature of Eq.~\eqref{eq:speckle_gen_sol} is that the modes $X_{0,\ell}$ and $\Pi_{0,\ell}^*$ are $\pi$-periodic functions of $\varphi(z)$.
Since $\varphi(z)$ contains a term linear in $z$, in general this leads to oscillations of specific observables in $z$. To illustrate this
phenomenon and as an application of the above formalism, we now consider a concrete scenario where a fluid of light is initially prepared in the form
of a (two-component) plane-wave background plus a small fluctuating field (see Fig.~\ref{fig:setup}):
\begin{equation}
\begin{bmatrix}
\mathcal{E}_+(\vec{r}_\perp,z=0) \\
\mathcal{E}_-(\vec{r}_\perp,z=0)
\end{bmatrix}
= \sqrt{I_0}
\begin{bmatrix}
\cos\frac{\vartheta_0}{2} + \epsilon \phi_+(\vec{r}_\perp) \\
\sin\frac{\vartheta_0}{2} + \epsilon \phi_-(\vec{r}_\perp)
\end{bmatrix},
\label{eq:speckle_in_f}
\end{equation}
where $0 < \epsilon \ll 1$, and $\phi_\alpha$ ($\alpha = \pm$) is a two-component random complex speckle field of two-point correlation
$\langle \phi_\alpha(\vec{r}_\perp) \phi_{\alpha'}^*(\vec{r}_\perp + \Delta \vec{r}_\perp)\rangle = \gamma(\Delta \vec{r}_\perp) \delta_{\alpha\alpha'}$,
the brackets denoting statistical averaging. For definiteness we consider a Gaussian correlation $\gamma(\Delta \vec{r}_\perp) = \exp(-|\Delta \vec{r}_\perp|^2
/ 4 \sigma^2)$, with $\sigma$ the correlation length. An initial state of the form of Eq.~\eqref{eq:speckle_in_f} was recently exploited
experimentally~\cite{Abuzarli2022} in order to realize an optical analog of the quench dynamics of thermal fluctuations in a 2D Bose gas,
following Ref.~\cite{BardonBrun2020}. The mode weights $C_\ell(\vec{q}_\perp)$ giving access to the state vector~\eqref{eq:speckle_gen_sol} at arbitrary $z$
are easily deduced from the $z = 0$ values of the fields [Eq.~\eqref{eq:speckle_in_f}] and the conjugate momenta, as shown in the Supplemental
Material~\cite{Supplemental}. The latter are fixed requiring the vanishing of the weights of backward-propagating modes, as shown in the Supplemental
Material~\cite{Supplemental}.
\begin{figure}
\includegraphics[scale=1.06]{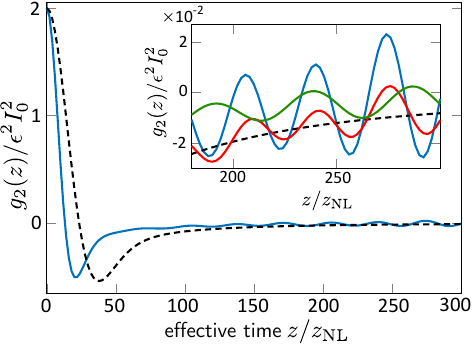}
\caption{Effective-time evolution of the intensity-intensity correlation
function. In the main plot we compare the exact (solid blue curve)
and paraxial (dashed curve) predictions for $\beta_0 \sigma = 15$
and the same parameters $\vartheta_0 = \pi/4$, $g_d I_0 / \beta_0 = 0.2$,
and $g_s I_0 / \beta_0 = 0.05$ as in Fig.~\ref{fig:omega_ell}. $z$ is measured
in units of the nonlinear length $z_{\mathrm{NL}} = 1 / 2 g_d I_0$. In the
inset we enlarge the large-$z$ tail and include two additional curves
showing the results for $\beta_0 \sigma = 22$ (red curve) and
$g_s I_0 / \beta_0 = 0.035$ (green curve), the other parameters being
the same as above.}
\label{fig:g2_ell}
\end{figure}
The knowledge of the weights enables one to compute any statistical observable. One of the simplest is the function $g_2(z) \equiv
\langle \delta I(\vec{r}_\perp,0) \delta I(\vec{r}_\perp,z) \rangle$, which expresses how intensity fluctuations at a finite effective time $z > 0$
and at a given point in the transverse plane are correlated with their $z = 0$ value (see Fig.~\ref{fig:setup}). We find that~\cite{Supplemental}
\begin{equation}
\begin{split}
\frac{g_2(z)}{\epsilon^2 I_0^2} = \sum_\ell \int_0^{\infty} \!\frac{q_\perp\dif q_\perp}{2\pi} \, \tilde{\gamma}(q_\perp) K_\ell(q_\perp, z)
\me^{- \mi \Omega_\ell(q_\perp) z},
\end{split}
\label{eq:speckle_2corr}
\end{equation}
where $\tilde{\gamma}(q_\perp)$ is the speckle power spectrum. In the paraxial regime $q_\perp \to 0$ and $g_{d,s} I_0 \ll \beta_0$, the coefficients
for the density and the spin modes $K_{d(s)}\simeq1/2 \pm {(g_d + g_s \cos 2 \vartheta_0)}/{2\sqrt{g_d^2 + g_s^2 + 2 g_d g_s \cos 2\vartheta_0}}$
reduce to constants independent of $q_\perp$ and $z$. $g_2$ in the paraxial limit is shown in Fig.~\ref{fig:g2_ell} as a function of $z$ (dashed curve),
at fixed background polarization, nonlinearities, and correlation length. At $z=0$, $g_2(0)/\epsilon^2I_0^2 = 2$, which corresponds to the Rayleigh law
of the speckle field. At large $z$, the correlation is negative and approaches zero as $z \to + \infty$ following $g_2(z)/\epsilon^2 I_0^2 \simeq
- (\sqrt{g_d I_0/\beta_0} {z}/{2 \sigma})^{-2}$. In between these two asymptotic limits $g_2$ changes sign, in agreement with the sum rule
$\int_0^\infty \dif z \, g_2(z) = 0$.

Beyond the paraxial limit, the emerging Floquet structure of the spectrum gives rise to periodic $z$ oscillations of the coefficients $K_\ell$. This
manifests itself in oscillations of $g_2$ about its background value at large $z$, as seen in Fig.~\ref{fig:g2_ell} (solid blue curve). The magnitude
of these oscillations is directly controlled by two main physical parameters: first, the speckle correlation length $\sigma$, which controls by how much
the fluid of light deviates from paraxiality, a necessary condition for the system to exhibit SOC, and second, the coupling strength $g_s$, which gives
rise to nonlinear birefringence. These dependences are illustrated in the inset of Fig.~\ref{fig:g2_ell}: the oscillations' amplitude decreases (increases)
with $\beta_0\sigma$ ($g_s I_0/\beta_0$). In particular, at large $\sigma$ the spectrum $\tilde{\gamma}(q_\perp)$ becomes very narrow, so that the
integral [Eq.~\eqref{eq:speckle_2corr}] becomes dominated by the small-$q_\perp$ modes, and $K_\ell(q_\perp,z)$ can be replaced by its constant,
$q_\perp = 0$ value. The oscillation frequency, in contrast, is practically independent of $\sigma$. It is mainly governed by the nonlinear birefringence
mismatch $\Delta k$ of the fluid's two components. From a Fourier analysis of $g_2(z)$, we find that these oscillations are essentially harmonic at large
enough $\sigma$, with a frequency $\simeq 0.8 \, |\Delta k|$~\cite{Supplemental}.

\begin{figure}
\includegraphics[scale=0.67]{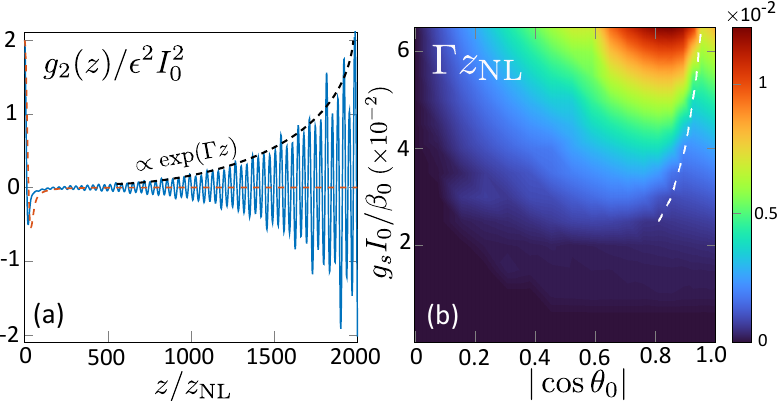}
\caption{(a) Evolution of $g_2(z)$ up to very large effective times $z$,
revealing a phenomenon of parametric instability with a growth rate
$\Gamma$. Parameters are the same as in Fig.~\ref{fig:g2_ell}.
(b) Growth rate as a function of the background polarization,
$\cos \vartheta_0$, and the spin-dependent nonlinear coupling,
$g_s I_0/\beta_0$, at fixed $g_d I_0 / \beta_0 = 0.2$ and
$\beta_0 \sigma = 15$. The white dashed curve pinpoints a jump
of $\Gamma$, corresponding to the appearance of an additional
imaginary frequency participating in the dynamics.}
\label{fig:g2_ell_growth}
\end{figure}

A second remarkable consequence of the Floquet structure of the fluid's spectrum is a phenomenon of parametric resonance~\cite{Landau_Lifshitz_01_book},
analogous to that observed in periodically driven atomic~\cite{Tozzo2005,Creffield2009,Bukov2015,Lellouch2017,Boulier2019} and photonic~\cite{
Sarchi2008,Gavrilov2017,Gavrilov2020,Carlon2020,Koniakhin2021} systems. Parametric resonances are associated with the spontaneous emergence of complex
frequencies in the Floquet spectrum, see Fig.~\ref{fig:omega_ell}(b). They can occur when the two lowest positive eigenvalues $\Omega^{(0)}_{1,2}$ of the
time average of the matrix entering Eq.~\eqref{eq:ham_eq} fulfill $\smash{\Omega^{(0)}_a(q_\perp) + \Omega^{(0)}_b(q_\perp) \simeq n \Delta k}$ for $a,b = 1,2$
and integer $n > 0$~\cite{Lellouch2017,Supplemental}. Parametric resonances induce an exponential growth of the population of the corresponding Bogoliubov
modes at large $z$, resulting in a strong increase of the fringe amplitude in $g_2$, as illustrated in Fig.~\ref{fig:g2_ell_growth}(a). To explore more
systematically the occurrence of these instabilities, we have analyzed their characteristic growth rate $\Gamma$ for various interaction strengths $g_s I_0$
and background polarizations $\cos\vartheta_0$.
To this aim, we studied the evolution of $g_2(z)$ within a fixed range of large values of $z$ and fitted the data with an exponential law of the form
$\exp(\Gamma z)$. The results are summarized by the diagram in Fig.~\ref{fig:g2_ell_growth}(b). As expected, $\Gamma$ vanishes when either $\cos\vartheta_0
\to 0$ (linearly polarized background) or $g_s \to 0$ (no spin-dependent nonlinearity). At large enough $g_s$, Fig.~\ref{fig:g2_ell_growth}(b) also reveals
a sharp increase of $\Gamma$ as $|\cos\vartheta_0|$ is decreased from $1$, pinpointed by the white dashed curve. This corresponds to the appearance of an
additional complex frequency that starts to dominate the exponential growth in the chosen effective-time window, our fitting procedure yielding the largest
$\Gamma$. Note that the exponential growth of $g_2$ is physical as long as $\langle \left( \delta I \right)^2 \rangle / I_0^2 \ll 1$ [and same for
$\langle \left( \delta \vartheta \right)^2 \rangle$ and $\langle \left( \delta \chi \right)^2 \rangle$], corresponding to $z \ll \ln (\epsilon^{-2}) /
\max_{\ell, q_\perp} [\Im \Omega_\ell(q_\perp)]$. At larger $z$, nonlinear effects not captured by the Bogoliubov approximation eventually cause a saturation
of the value of physical observables~\cite{Tozzo2005,Lellouch2017}.

Several experimental studies of the polarization dependence of the nonlinear refractive index are available~\cite{Maker1964,Mayer1964,
Maker1965,Duguay1969,Owyoung1972,Duguay1976,Hellwarth1977,PhuXuan1978,Shen1984,Boyd2003}. Typically one finds $g_s / g_d$ of order $10^{-1}$, not far from the
value $g_s / g_d = 0.25$ chosen in this work. Taking this value and a nonlinear strength $g_d I_0 / \beta_0 \sim 10^{-3}$~\cite{Abuzarli2022}, we find that the
oscillations of $g_2(z)$ are visible at propagation distances $z$ of the order of a few tens of meters. Such a scale points toward optical fibers (see, e.g.,
Ref.~\cite{Vanderhaegen2022}) as good candidates for the observation of the phenomena predicted in this Letter.

\begin{acknowledgments}
We thank D. Ballarini, T. Bienaim\'{e}, I. Carusotto, D. Delande, L. Dominici, Q. Glorieux, N. Pavloff, K. Sacha, and J. Zakrzewski for fruitful
discussions.  We acknowledge financial support from the Agence Nationale de la Recherche (Grant No. ANR-19-CE30-0028-01 CONFOCAL) and from the Italian
Ministry of University and Research (MUR) through the PRIN project INPhoPOL (Grant No. 2017P9FJBS) and the PNRR MUR project PE0000023 - NQSTI.
\end{acknowledgments}

\end{document}


\title{Supplemental Material:\texorpdfstring{\\}{} Time Translation Symmetry Breaking in an Isolated Spin-Orbit-Coupled Fluid of Light}

\author{Giovanni I. Martone}
\email{giovanni\textunderscore italo.martone@lkb.upmc.fr}
\affiliation{Laboratoire Kastler Brossel,
Sorbonne Universit\'{e}, CNRS, ENS-PSL Research University, 
Coll\`{e}ge de France; 4 Place Jussieu, 75005 Paris, France}
\affiliation{CNR NANOTEC, Institute of Nanotechnology,
Via Monteroni, 73100 Lecce, Italy}
\affiliation{INFN, Sezione di Lecce, 73100 Lecce, Italy}

\author{Nicolas Cherroret}
\email{nicolas.cherroret@lkb.upmc.fr}
\affiliation{Laboratoire Kastler Brossel,
Sorbonne Universit\'{e}, CNRS, ENS-PSL Research University, 
Coll\`{e}ge de France; 4 Place Jussieu, 75005 Paris, France}

\begin{abstract}
In this Supplemental Material we provide further details about the calculations leading to the results of the main text, in particular the nonparaxial
Bogoliubov action (Sec.~\ref{sec:nonpar_bogo_act}), the paraxial limit and the sound velocity (Sec.~\ref{sec:paraxial_lim}), the solution of the
Bogoliubov equations (Sec.~\ref{sec:bogo_hill}), the parametric resonance conditions (Sec.~\ref{sec:param_reson}), the choice of the input conjugate
momenta (Sec.~\ref{sec:input_conj_mom}), and the expression of the intensity-intensity correlation function (Sec.~\ref{sec:g2}).
\end{abstract}

\maketitle

\section{Nonparaxial Bogoliubov action}
\label{sec:nonpar_bogo_act}
The starting point for the formulation of the Bogoliubov theory for a nonparaxial fluid of light is represented by its Lagrangian density
[Eq.~(2) of the main text]. First, one has to rewrite this quantity in terms of the density-phase variables. The resulting expression, that can
be found in Ref.~\cite{Martone2021}, should then be expanded up to second order in the small fluctuations about the background field. After switching
to the Fourier space, it is convenient to decompose the quadratic Lagrangian as $\tilde{\mathcal{L}}^{(2)} = \tilde{\mathcal{L}}_z^{(2)}
+ \tilde{\mathcal{L}}_\perp^{(2)}$, where $\tilde{\mathcal{L}}_z^{(2)}$ is the term that contains the longitudinal field $\tilde{\mathcal{E}}_z$,
while $\tilde{\mathcal{L}}_\perp^{(2)}$ depends only on the transverse components. In the case of the elliptically polarized background considered in
the present work one has
\begin{equation}
\begin{split}
\tilde{\mathcal{L}}_z^{(2)} = {}&{} - \frac{1}{2\beta_0}
\left[ (q_\perp^2 - k_0^2) \tilde{\mathcal{E}}_z^*(\vec{q}_\perp,z) \tilde{\mathcal{E}}_z(\vec{q}_\perp,z)
+ \frac{\Delta k_0^2}{2} \tilde{\mathcal{E}}_z(\vec{q}_\perp,z) \tilde{\mathcal{E}}_z(-\vec{q}_\perp,z)
+ \frac{\Delta k_0^2}{2} \tilde{\mathcal{E}}_z^*(\vec{q}_\perp,z) \tilde{\mathcal{E}}_z^*(-\vec{q}_\perp,z) \right] \\
&{} + \frac{\mi q_\perp}{2\beta_0}\sqrt{\frac{I_0}{2}} \left[ \mathcal{A}(\vec{q}_\perp,z) \tilde{\mathcal{E}}_z^*(\vec{q}_\perp,z)
- \mathcal{A}^*(\vec{q}_\perp,z) \tilde{\mathcal{E}}_z(\vec{q}_\perp,z) \right] \, ,
\end{split}
\label{eq:lagr_z_2}
\end{equation}
where $k_0^2 = \beta_0^2 - 2 \beta_0 (g_d + g_s) I_0$, $\Delta k_0^2 = 2 \beta_0 g_s I_0 \sin\vartheta_0$, and
\begin{equation}
\begin{split}
\mathcal{A}(\vec{q}_\perp,z) =  {}&{}
\left( \cos\frac{\vartheta_0}{2} \, \me^{\mi \varphi} - \sin\frac{\vartheta_0}{2} \, \me^{- \mi \varphi} \right)
\frac{\delta\dot{\tilde{I}}(\vec{q}_\perp,z)}{2 I_0}
- \left( \sin\frac{\vartheta_0}{2} \, \me^{\mi \varphi} + \cos\frac{\vartheta_0}{2} \, \me^{- \mi \varphi} \right)
\frac{\delta\dot{\tilde{\vartheta}}(\vec{q}_\perp,z)}{2} \\
{}&{} + \mi \left( \cos\frac{\vartheta_0}{2} \, \me^{\mi \varphi} - \sin\frac{\vartheta_0}{2} \, \me^{- \mi \varphi} \right)
\dot{\tilde{\Theta}}(\vec{q}_\perp,z)
+ \mi \left( \cos\frac{\vartheta_0}{2} \, \me^{\mi \varphi} + \sin\frac{\vartheta_0}{2} \, \me^{- \mi \varphi} \right)
\frac{\delta\dot{\tilde{\chi}}(\vec{q}_\perp,z)}{2} \\
{}&{} + \mi \left( k_+ \cos\frac{\vartheta_0}{2} \, \me^{\mi \varphi} - k_- \sin\frac{\vartheta_0}{2} \, \me^{- \mi \varphi} \right)
\frac{\delta\tilde{I}(\vec{q}_\perp,z)}{2 I_0}
- \mi \left( k_+ \sin\frac{\vartheta_0}{2} \, \me^{\mi \varphi} + k_- \cos\frac{\vartheta_0}{2} \, \me^{- \mi \varphi} \right)
\frac{\delta\tilde{\vartheta}(\vec{q}_\perp,z)}{2} \\
{}&{} - \left( k_+ \cos\frac{\vartheta_0}{2} \, \me^{\mi \varphi} - k_- \sin\frac{\vartheta_0}{2} \, \me^{- \mi \varphi} \right)
\tilde{\Theta}(\vec{q}_\perp,z)
- \left( k_+ \cos\frac{\vartheta_0}{2} \, \me^{\mi \varphi} + k_- \sin\frac{\vartheta_0}{2} \, \me^{- \mi \varphi} \right)
\frac{\delta\tilde{\chi}(\vec{q}_\perp,z)}{2} \, .
\end{split}
\label{eq:lagr_z_2_coeff}
\end{equation}
We recall that $\varphi = \varphi_q + \Delta k \;\!\!\! z / 2$. Since the Bogoliubov Lagrangian density and the corresponding
action $\mathcal{S}^{(2)} = \int \dif z \int {\dif^2 \vec{q}_\perp}/{(2\pi)^2} \tilde{\mathcal{L}}^{(2)}$ do not depend on the effective-time
derivatives of $\tilde{\mathcal{E}}_z$, the Euler-Lagrange equation $\delta \mathcal{S}^{(2)} / \delta \tilde{\mathcal{E}}_z^*(\vec{q}_\perp) = 0$ and
its complex conjugate yield the relation 
\begin{equation}
\tilde{\mathcal{E}}_z(\vec{q}_\perp,z) = \sqrt{\frac{I_0}{2}}
\frac{\mi q_\perp \left[(q_\perp^2 - k_0^2) \mathcal{A}(\vec{q}_\perp,z)
+ \Delta k_0^2 \mathcal{A}^*(-\vec{q}_\perp,z) \right]}{(q_\perp^2 - k_0^2)^2 - (\Delta k_0^2)^2} \, ,
\label{eq:Ez}
\end{equation}
which allows us to eliminate the longitudinal field from the Lagrangian in favor of the variables related to the transverse components and their
derivatives. The final result corresponds to Eq.~(4) of the main text, i.e.,
\begin{equation}
\tilde{\mathcal{L}}^{(2)} = \dot{X}^\dagger \Lambda_2 \dot{X} + \dot{X}^\dagger \Lambda_1 X
+ X^\dagger \Lambda_1^T \dot{X} - X^\dagger \Lambda_0 X \, ,
\label{eq:lagr_2} 
\end{equation}
where we have introduced three $4 \times 4$ matrices having the structure
\begin{equation}
\Lambda_k = -\frac{I_0}{2 \beta_0}
\begin{bmatrix}
(\Lambda_k)_{1,1} & (\Lambda_k)_{1,2} & (\Lambda_k)_{1,3} & (\Lambda_k)_{1,4} \\
(\Lambda_k)_{2,1} & (\Lambda_k)_{2,2} & (\Lambda_k)_{2,3} & (\Lambda_k)_{2,4} \\
(\Lambda_k)_{3,1} & (\Lambda_k)_{3,2} & (\Lambda_k)_{3,3} & (\Lambda_k)_{3,4} \\
(\Lambda_k)_{4,1} & (\Lambda_k)_{4,2} & (\Lambda_k)_{4,3} & (\Lambda_k)_{4,4} \\
\end{bmatrix}
\qquad (k=0,1,2) \, .
\label{eq:Lambda_mat}
\end{equation}
We now give the expressions of all the entries of these matrices. We first define
\begin{subequations}
\label{eq:Lambda_mat_coeff}
\begin{align}
\qqa {}&{} = \frac{q_\perp^2}{2} \frac{q_\perp^2 - k_0^2}{(q_\perp^2 - k_0^2)^2 - (\Delta k_0^2)^2} \, ,
\label{eq:Lambda_mat_coeff_1} \\
\qqb {}&{} = \frac{q_\perp^2}{2} \frac{\Delta k_0^2}{(q_\perp^2 - k_0^2)^2 - (\Delta k_0^2)^2} \, .
\label{eq:Lambda_mat_coeff_2}
\end{align}
\end{subequations}
Then, the entries of $\Lambda_0$ are
\begin{subequations}
\label{eq:Lambda_0}
\begin{align}
\begin{split}
&{} (\Lambda_0)_{1,1} =
\begin{aligned}[t]
{}&{} - \frac{q_\perp^2}{2} - 4 \beta_0 (g_d + g_s \cos^2 \vartheta_0) I_0
+ \qqa \left(k_+^2 \cos^2 \frac{\vartheta_0}{2} + k_-^2 \sin^2 \frac{\vartheta_0}{2} \right) - \qqb k_+ k_- \sin\vartheta_0 \\
{}&{} - \left[ \frac{q_\perp^2}{2} \sin\vartheta_0
+ \qqa k_+ k_- \sin\vartheta_0 - \qqb \left(k_+^2 \cos^2 \frac{\vartheta_0}{2} + k_-^2 \sin^2 \frac{\vartheta_0}{2}\right) \right] \cos 2\varphi \, ,
\end{aligned}
\end{split}
\label{eq:Lambda_0_11} \\
\begin{split}
&{} (\Lambda_0)_{1,2} = (\Lambda_0)_{2,1} =
\begin{aligned}[t]
{}&{} 2 \beta_0 g_s I_0 \sin 2\vartheta_0 - \qqa \frac{k_+^2 - k_-^2}{2} \sin\vartheta_0 - \qqb k_+ k_- \cos\vartheta_0 \\
{}&{} - \left[ \frac{q_\perp^2}{2} \cos\vartheta_0 + \qqa k_+ k_- \cos\vartheta_0 + \qqb \frac{k_+^2 - k_-^2}{2} \sin\vartheta_0 \right]
\cos 2\varphi \, ,
\end{aligned}
\end{split}
\label{eq:Lambda_0_12} \\
&{} (\Lambda_0)_{1,3} = (\Lambda_0)_{3,1} =
- \qqb \left(k_+^2 \cos^2 \frac{\vartheta_0}{2} - k_-^2 \sin^2 \frac{\vartheta_0}{2}\right) \sin 2\varphi \, ,
\label{eq:Lambda_0_13} \\
&{} (\Lambda_0)_{1,4} = (\Lambda_0)_{4,1} =
\left[ \frac{q_\perp^2}{2} \sin\vartheta_0 + \qqa k_+ k_- \sin\vartheta_0
- \qqb \left(k_+^2 \cos^2 \frac{\vartheta_0}{2} + k_-^2 \sin^2 \frac{\vartheta_0}{2}\right) \right] \sin 2\varphi \, ,
\label{eq:Lambda_0_14} \\
\begin{split}
&{} (\Lambda_0)_{2,2} =
\begin{aligned}[t]
{}&{} - \frac{q_\perp^2}{2} - 4 \beta_0 g_s I_0 \sin^2 \vartheta_0
+ \qqa \left(k_+^2 \sin^2 \frac{\vartheta_0}{2} + k_-^2 \cos^2 \frac{\vartheta_0}{2} \right) + \qqb k_+ k_- \sin\vartheta_0 \\
{}&{} + \left[ \frac{q_\perp^2}{2} \sin\vartheta_0 + \qqa k_+ k_- \sin\vartheta_0
+ \qqb \left(k_+^2 \sin^2 \frac{\vartheta_0}{2} + k_-^2 \cos^2 \frac{\vartheta_0}{2}\right) \right] \cos 2\varphi  \, ,
\end{aligned}
\end{split}
\label{eq:Lambda_0_22} \\
&{} (\Lambda_0)_{2,3} = (\Lambda_0)_{3,2} =
\left[ \frac{q_\perp^2}{2} + \qqa k_+ k_- + \qqb \frac{k_+^2 + k_-^2}{2} \sin\vartheta_0 \right] \sin 2\varphi \, ,
\label{eq:Lambda_0_23} \\
&{} (\Lambda_0)_{2,4} = (\Lambda_0)_{4,2} =
\left[ \frac{q_\perp^2}{2} \cos\vartheta_0
+ \qqa \, k_+ k_- \cos\vartheta_0 + \qqb \frac{k_+^2 - k_-^2}{2}  \sin\vartheta_0 \right] \sin 2\varphi  \, ,
\label{eq:Lambda_0_24} \\
\begin{split}
&{} (\Lambda_0)_{3,3} =
\begin{aligned}[t]
{}&{} - \frac{q_\perp^2}{2}
+ \qqa \left(k_+^2 \cos^2 \frac{\vartheta_0}{2} + k_-^2 \sin^2 \frac{\vartheta_0}{2}\right) + \qqb k_+ k_- \sin\vartheta_0 \\
{}&{} - \left[ \frac{q_\perp^2}{2} \sin\vartheta_0
+ \qqa k_+ k_- \sin\vartheta_0 + \qqb \left(k_+^2 \cos^2 \frac{\vartheta_0}{2} + k_-^2 \sin^2 \frac{\vartheta_0}{2}\right) \right] \cos 2\varphi \, ,
\end{aligned}
\end{split}
\label{eq:Lambda_0_33} \\
\begin{split}
&{} (\Lambda_0)_{3,4} = (\Lambda_0)_{4,3} =
\begin{aligned}[t]
{}&{} - \frac{q_\perp^2}{2} \cos\vartheta_0
+ \qqa \left(k_+^2 \cos^2 \frac{\vartheta_0}{2} - k_-^2 \sin^2 \frac{\vartheta_0}{2}\right) \\
{}&{} - \qqb \left(k_+^2 \cos^2 \frac{\vartheta_0}{2} - k_-^2 \sin^2 \frac{\vartheta_0}{2}\right) \cos 2\varphi \, ,
\end{aligned}
\end{split}
\label{eq:Lambda_0_34} \\
\begin{split}
&{} (\Lambda_0)_{4,4} =
\begin{aligned}[t]
{}&{} - \frac{q_\perp^2}{2}
+ \qqa \left(k_+^2 \cos^2 \frac{\vartheta_0}{2} + k_-^2 \sin^2 \frac{\vartheta_0}{2}\right) - \qqb k_+ k_- \sin\vartheta_0 \\
{}&{} + \left[ \frac{q_\perp^2}{2} \sin\vartheta_0 + \qqa k_+ k_- \sin\vartheta_0
- \qqb \left(k_+^2 \cos^2 \frac{\vartheta_0}{2} + k_-^2 \sin^2 \frac{\vartheta_0}{2}\right) \right] \cos 2\varphi \, .
\end{aligned}
\end{split}
\label{eq:Lambda_0_44}
\end{align}
\end{subequations}
The entries of $\Lambda_1$ are
\begin{subequations}
\label{eq:Lambda_1}
\begin{align}
&{} (\Lambda_1)_{1,1} =
- \left[ \qqa \frac{k_+ - k_-}{2} \sin\vartheta_0
+ \qqb \left( k_+ \cos^2\frac{\vartheta_0}{2} - k_- \sin^2\frac{\vartheta_0}{2} \right) \right] \sin 2\varphi \, ,
\label{eq:Lambda_1_11} \\
&{} (\Lambda_1)_{1,2} =
\left[ \qqa \left( k_+ \sin^2\frac{\vartheta_0}{2} + k_- \cos^2\frac{\vartheta_0}{2} \right)
+ \qqb \frac{k_+ + k_-}{2} \sin\vartheta_0 \right] \sin 2\varphi \, ,
\label{eq:Lambda_1_12} \\
\begin{split}
&{} (\Lambda_1)_{1,3} =
\begin{aligned}[t]
{}&{} \qqa \left(k_+ \cos^2 \frac{\vartheta_0}{2} + k_- \sin^2 \frac{\vartheta_0}{2}\right)
+ \qqb \frac{k_+ + k_-}{2} \sin\vartheta_0 \\
{}&{} - \left[ \qqa \frac{k_+ + k_-}{2} \sin\vartheta_0
+ \qqb \left(k_+ \cos^2 \frac{\vartheta_0}{2} + k_- \sin^2 \frac{\vartheta_0}{2}\right) \right] \cos 2\varphi \, ,
\end{aligned}
\end{split}
\label{eq:Lambda_1_13} \\
\begin{split}
&{} (\Lambda_1)_{1,4} =
\begin{aligned}[t]
{}&{} \qqa \left(k_+ \cos^2 \frac{\vartheta_0}{2} - k_- \sin^2 \frac{\vartheta_0}{2}\right)
+ \qqb \frac{k_+ - k_-}{2} \sin\vartheta_0 \\
{}&{} - \left[ \qqa \frac{k_+ - k_-}{2} \sin\vartheta_0
+ \qqb \left(k_+ \cos^2 \frac{\vartheta_0}{2} - k_- \sin^2 \frac{\vartheta_0}{2}\right) \right] \cos 2\varphi \, ,
\end{aligned}
\end{split}
\label{eq:Lambda_1_14} \\
&{} (\Lambda_1)_{2,1} =
\left[ - \qqa \left( k_+ \cos^2\frac{\vartheta_0}{2} + k_- \sin^2\frac{\vartheta_0}{2} \right)
+ \qqb \frac{k_+ + k_-}{2} \sin\vartheta_0 \right] \sin 2\varphi \, ,
\label{eq:Lambda_1_21} \\
&{} (\Lambda_1)_{2,2} =
\left[ \qqa \frac{k_+ - k_-}{2} \sin\vartheta_0
- \qqb \left( k_+ \sin^2\frac{\vartheta_0}{2} - k_- \cos^2\frac{\vartheta_0}{2} \right) \right] \sin 2\varphi \, ,
\label{eq:Lambda_1_22} \\
\begin{split}
&{} (\Lambda_1)_{2,3} =
\begin{aligned}[t]
{}&{} - \qqa \frac{k_+ - k_-}{2} \sin\vartheta_0
+ \qqb \left(k_+ \cos^2 \frac{\vartheta_0}{2} - k_- \sin^2 \frac{\vartheta_0}{2}\right) \\
{}&{} - \left[\qqa \left(k_+ \cos^2 \frac{\vartheta_0}{2} - k_- \sin^2 \frac{\vartheta_0}{2}\right)
- \qqb \frac{k_+ - k_-}{2} \sin\vartheta_0\right] \cos 2\varphi \, ,
\end{aligned}
\end{split}
\label{eq:Lambda_1_23} \\
\begin{split}
&{} (\Lambda_1)_{2,4} =
\begin{aligned}[t]
{}&{} - \qqa \frac{k_+ + k_-}{2} \sin\vartheta_0
+ \qqb \left(k_+ \cos^2 \frac{\vartheta_0}{2} + k_- \sin^2 \frac{\vartheta_0}{2}\right) \\
{}&{} - \left[\qqa \left(k_+ \cos^2 \frac{\vartheta_0}{2} + k_- \sin^2 \frac{\vartheta_0}{2}\right)
- \qqb \frac{k_+ + k_-}{2} \sin\vartheta_0 \right] \cos 2\varphi \, ,
\end{aligned}
\end{split}
\label{eq:Lambda_1_24} \\
\begin{split}
&{} (\Lambda_1)_{3,1} =
\begin{aligned}[t]
{}&{} \left[2 - \qqa\right] \left(k_+ \cos^2\frac{\vartheta_0}{2} + k_- \sin^2\frac{\vartheta_0}{2}\right)
+ \qqb \frac{k_+ + k_-}{2} \sin\vartheta_0 \\
{}&{} + \left[\qqa \frac{k_+ + k_-}{2} \sin\vartheta_0
- \qqb \left(k_+ \cos^2 \frac{\vartheta_0}{2} + k_- \sin^2 \frac{\vartheta_0}{2}\right)\right] \cos 2\varphi \, ,
\end{aligned}
\end{split}
\label{eq:Lambda_1_31} \\
\begin{split}
&{} (\Lambda_1)_{3,2} =
\begin{aligned}[t]
{}&{} - \left[2 - \qqa\right] \frac{k_+ - k_-}{2} \sin\vartheta_0
- \qqb \left(k_+ \sin^2 \frac{\vartheta_0}{2} - k_- \cos^2 \frac{\vartheta_0}{2}\right) \\
{}&{} - \left[ \qqa \left(k_+ \sin^2 \frac{\vartheta_0}{2} - k_- \cos^2 \frac{\vartheta_0}{2}\right)
- \qqb \frac{k_+ - k_-}{2} \sin\vartheta_0 \right] \cos 2\varphi \, ,
\end{aligned}
\end{split}
\label{eq:Lambda_1_32} \\
&{} (\Lambda_1)_{3,3} =
\left[ - \qqa \frac{k_+ - k_-}{2} \sin\vartheta_0
+ \qqb \left( k_+ \cos^2\frac{\vartheta_0}{2} - k_- \sin^2\frac{\vartheta_0}{2} \right) \right] \sin 2\varphi \, ,
\label{eq:Lambda_1_33} \\
&{} (\Lambda_1)_{3,4} =
\left[ - \qqa \frac{k_+ + k_-}{2} \sin\vartheta_0
+ \qqb \left( k_+ \cos^2\frac{\vartheta_0}{2} + k_- \sin^2\frac{\vartheta_0}{2} \right) \right] \sin 2\varphi \, ,
\label{eq:Lambda_1_34} \\
\begin{split}
&{} (\Lambda_1)_{4,1} =
\begin{aligned}[t]
{}&{} \left[2 - \qqa\right]  \left(k_+ \cos^2\frac{\vartheta_0}{2} - k_- \sin^2\frac{\vartheta_0}{2}\right)
- \qqb \frac{k_+ - k_-}{2} \sin\vartheta_0 \\
{}&{} - \left[\qqa \frac{k_+ - k_-}{2} \sin\vartheta_0
+ \qqb \left(k_+ \cos^2 \frac{\vartheta_0}{2} - k_- \sin^2 \frac{\vartheta_0}{2}\right) \right] \cos 2\varphi \, ,
\end{aligned}
\end{split}
\label{eq:Lambda_1_41} \\
\begin{split}
&{} (\Lambda_1)_{4,2} =
\begin{aligned}[t]
{}&{} - \left[2 - \qqa\right] \frac{k_+ + k_-}{2} \sin\vartheta_0
+ \qqb \left(k_+ \sin^2 \frac{\vartheta_0}{2} + k_- \cos^2 \frac{\vartheta_0}{2}\right) \\
{}&{} + \left[ \qqa \left(k_+ \sin^2 \frac{\vartheta_0}{2} + k_- \cos^2 \frac{\vartheta_0}{2}\right)
+ \qqb \frac{k_+ + k_-}{2} \sin\vartheta_0 \right] \cos 2\varphi \, ,
\end{aligned}
\end{split}
\label{eq:Lambda_1_42} \\
&{} (\Lambda_1)_{4,3} =
\left[ \qqa \frac{k_+ + k_-}{2} \sin\vartheta_0
+ \qqb \left( k_+ \cos^2\frac{\vartheta_0}{2} + k_- \sin^2\frac{\vartheta_0}{2} \right) \right] \sin 2\varphi \, ,
\label{eq:Lambda_1_43} \\
&{} (\Lambda_1)_{4,4} =
\left[ \qqa \frac{k_+ - k_-}{2} \sin\vartheta_0
+ \qqb \left( k_+ \cos^2\frac{\vartheta_0}{2} - k_- \sin^2\frac{\vartheta_0}{2} \right) \right] \sin 2\varphi \, .
\label{eq:Lambda_1_44}
\end{align}
\end{subequations}
Finally, the entries of $\Lambda_2$ are
\begin{subequations}
\label{eq:Lambda_2}
\begin{align}
&{} (\Lambda_2)_{1,1} =
1 - \qqa - \qqb \sin\vartheta_0 + \left[\qqa \sin\vartheta_0 + \qqb\right] \cos 2\varphi \, ,
\label{eq:Lambda_2_11} \\
&{} (\Lambda_2)_{1,2} = (\Lambda_2)_{2,1} =
- \qqb \cos\vartheta_0 + \qqa \cos\vartheta_0 \cos 2\varphi \, ,
\label{eq:Lambda_2_12} \\
&{} (\Lambda_2)_{1,3} = (\Lambda_2)_{3,1} =
- \qqb \cos\vartheta_0 \sin 2\varphi \, ,
\label{eq:Lambda_2_13} \\
&{} (\Lambda_2)_{1,4} = (\Lambda_2)_{4,1} =
- \left[\qqa \sin\vartheta_0 + \qqb\right] \sin 2\varphi \, ,
\label{eq:Lambda_2_14} \\
&{} (\Lambda_2)_{2,2} =
1 - \qqa + \qqb \sin\vartheta_0 - \left[\qqa \sin\vartheta_0 - \qqb\right] \cos 2\varphi \, ,
\label{eq:Lambda_2_22} \\
&{} (\Lambda_2)_{2,3} = (\Lambda_2)_{3,2} =
- \left[\qqa - \qqb \sin\vartheta_0\right] \sin 2\varphi \, ,
\label{eq:Lambda_2_23} \\
&{} (\Lambda_2)_{2,4} = (\Lambda_2)_{4,2} =
- \qqa \cos\vartheta_0 \sin 2\varphi \, ,
\label{eq:Lambda_2_24} \\
&{} (\Lambda_2)_{3,3} =
1 - \qqa + \qqb \sin\vartheta_0 + \left[\qqa \sin\vartheta_0 - \qqb\right] \cos 2\varphi \, ,
\label{eq:Lambda_2_33} \\
&{} (\Lambda_2)_{3,4} = (\Lambda_2)_{4,3} =
\left[1 - \qqa\right] \cos\vartheta_0 - \qqb \cos\vartheta_0 \cos 2\varphi \, ,
\label{eq:Lambda_2_34} \\
&{} (\Lambda_2)_{4,4} =
1 - \qqa - \qqb \sin\vartheta_0 - \left[\qqa \sin\vartheta_0 + \qqb\right] \cos 2\varphi \, .
\label{eq:Lambda_2_44}
\end{align}
\end{subequations}

\section{Paraxial limit and sound velocities}
\label{sec:paraxial_lim}
\begin{figure}
\includegraphics[scale=1]{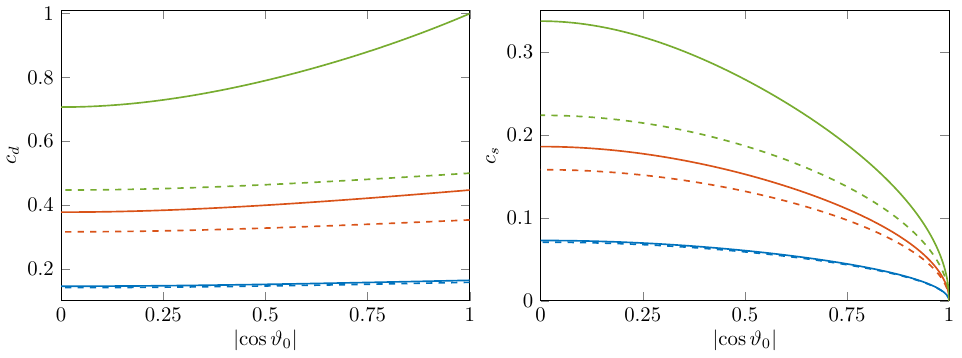}
\caption{Velocity of (left) density and (right) spin sound waves as a function
of the background polarization quantified by $\cos\vartheta_0$. The solid
curves represent the results of the full nonparaxial theory, while the dashed
ones correspond to the paraxial predictions~\eqref{eq:paraxial_sound}. We take
$g_d I_0 / \beta_0 = 0.02$ (blue lines), $0.1$ (red lines), $0.2$ (green lines),
with the same ratio $g_s / g_d = 0.25$ for all the curves.}
\label{fig:sound_vel}
\end{figure}

As stated in the main text, the paraxial Bogoliubov Lagrangian can be deduced by expanding the nonparaxial one [Eq.~\eqref{eq:lagr_2}] up to first
order in $\dot{X}/\beta_0$, $(q_\perp/\beta_0)^2$, and $g_{d,s}I_0 / \beta_0$. The final result is
\begin{equation}
\tilde{\mathcal{L}}_\mathrm{par}^{(2)}
= \dot{X}^\dagger \Lambda_{\mathrm{par},1} X + X^\dagger \Lambda_{\mathrm{par},1}^T \dot{X} - X^\dagger \Lambda_{\mathrm{par},0} X \, ,
\label{eq:paraxial_lagr} 
\end{equation}
where
\begin{equation}
\begin{gathered}
\Lambda_{\mathrm{par},1} = I_0
\begin{bmatrix}
0 & 0 & 0 & 0 \\
0 & 0 & 0 & 0 \\
-1 & 0 & 0 & 0 \\
-\cos\vartheta_0 & \sin\vartheta_0 & 0 & 0 \\
\end{bmatrix}
\, , \\
\Lambda_{\mathrm{par},0} = I_0
\begin{bmatrix}
\frac{q_\perp^2}{2\beta_0} + 2(g_d + g_s\cos^2\vartheta_0)I_0 & - g_s I_0 \sin 2\vartheta_0 & 0 & 0 \\
- g_s I_0 \sin 2\vartheta_0 & \frac{q_\perp^2}{2\beta_0} + 2 g_s I_0 \sin^2\vartheta_0 & 0 & 0 \\
0 & 0 & \frac{q_\perp^2}{2\beta_0} & \frac{q_\perp^2}{2\beta_0} \cos\vartheta_0 \\
0 & 0 & \frac{q_\perp^2}{2\beta_0} \cos\vartheta_0 & \frac{q_\perp^2}{2\beta_0}
\end{bmatrix} \, .
\end{gathered}
\label{eq:paraxial_lagr_mat}
\end{equation}
The Euler-Lagrange equation for $X$ takes the simple form
\begin{equation}
\dot{X} = - \left[(\Lambda_{\mathrm{par},1} - \Lambda_{\mathrm{par},1}^T)\right]^{-1} \Lambda_{\mathrm{par},0} X \, .
\label{eq:paraxial_el_eq}
\end{equation}
By making use of the Ansatz $X(\vec{q}_\perp,z) = X_0(\vec{q}_\perp) \me^{- \mi \Omega(\vec{q}_\perp) z}$, one can reduce this equation to a
four-dimensional eigenvalue problem. One finds four solutions characterized by the oscillation frequencies $\pm \Omega_d$ and $\pm \Omega_s$,
which exhibit the standard Bogoliubov form
\begin{equation}
\Omega_{d(s)}(q_\perp) = \sqrt{\frac{q_\perp^2}{2 \beta_0} \left[\frac{q_\perp^2}{2 \beta_0} + 2 \beta_0 c_{d(s)}^2\right]} \, .
\label{eq:paraxial_bogo_freq}
\end{equation}
These are the counterparts of the density and spin mode of a binary mixture of atomic Bose-Einstein condensates, featuring in-phase
and out-of-phase oscillations of the densities of the two spin components, respectively. The corresponding sound velocities are given by
\begin{equation}
c_{d(s)}^2 = \frac{(g_d + g_s \pm \sqrt{g_d^2 + g_s^2 + 2 g_d g_s \cos 2\vartheta_0}) I_0}{2\beta_0} \, ,
\label{eq:paraxial_sound}
\end{equation}
where the upper (lower) sign refers to the density (spin) mode, respectively.
In Fig.~\ref{fig:sound_vel} we plot the velocity of (left) density and (right) spin sound waves as a function of the background polarization.
The paraxial prediction~\eqref{eq:paraxial_sound} (dashed curves) is compared with the values obtained in the nonparaxial description of this work
(solid curves) by numerically computing the slope of the linear bands appearing in the low-$q_\perp$ part of the Bogoliubov spectrum (see
Fig.~2 of the main text). We consider several values of the nonlinear coupling $g_d I_0 / \beta_0$, keeping the ratio $g_s / g_d$ fixed. We notice
that the qualitative behavior does not change between the two frameworks. On the other hand, the quantitative discrepancy is negligible at small
$g_d I_0 / \beta_0$, but it grows significantly at increasing nonlinearity. We further mention that $c_d^2$ approaches the value
$g_d I_0 / (\beta_0 - 3 g_d I_0)$ predicted in Ref.~\cite{Martone2021} in the $\cos\vartheta_0 \to 0$ limit (linearly polarized background).
Concerning the spin sound velocity, we first recall that its value in the case of a linearly polarized background field was computed in
Ref.~\cite{Martone2021} and found to be anisotropic, i.e., depending on $\varphi$. Here we checked that $c_s^2 \to
g_s I_0 [\beta_0 - 2 (g_d + g_s) I_0] / \{[\beta_0 - (2 g_d + g_s) I_0] [\beta_0 - 2 (g_d  + 2 g_s) I_0]\}$ as $\cos\vartheta_0 \to 0$, which is the
prediction of Ref.~\cite{Martone2021} [see Eq.~(47) therein] evaluated at $\varphi = \pi / 4$. In this respect we point out that, unlike the frequency
spectrum of a linearly polarized background field studied in Ref.~\cite{Martone2021}, the quasifrequency spectrum computed in the present work is
independent of $\varphi_q$, that is, it is isotropic. However, it should be noticed that frequency and quasifrequency have different definitions
that only make sense in the $\Delta k = 0$ and $\Delta k \neq 0$ case, respectively; hence, it is not inconsistent that the two spectra have
qualitatively different features.

\section{Solution of Bogoliubov equations. Hill's method}
\label{sec:bogo_hill}
Let us rewrite Eq.~(5) of the main text as
\begin{equation}
\mi
\begin{bmatrix}
\dot{X} \\
\dot{\Pi}^*
\end{bmatrix}	
= 
\mathcal{B}
\begin{bmatrix}
X \\
\Pi^*
\end{bmatrix} \, ,
\label{eq:bogo_eq}
\end{equation}
where the $8 \times 8$ matrix
\begin{equation}
\mathcal{B} =
\mi
\begin{bmatrix}
- \Lambda_2^{-1} \Lambda_1 & \Lambda_2^{-1} \\
- (\Lambda_1^T \Lambda_2^{-1} \Lambda_1 + \Lambda_0) & (\Lambda_2^{-1} \Lambda_1)^T
\end{bmatrix}
\label{eq:bogo_mat}
\end{equation}
is a $\pi$-periodic function of $\varphi = \varphi_q + \Delta k \;\! z / 2$. As discussed in the main text, the Floquet theorem allows us
to write the
general integral of Eq.~\eqref{eq:bogo_eq} as a linear combination of eight independent solutions of the form
\begin{equation}
\begin{bmatrix}
X_\ell(\vec{q}_\perp,z) \\
\Pi_\ell^*(\vec{q}_\perp,z)
\end{bmatrix}
= \me^{- \mi \Omega_\ell(q_\perp) z}
\begin{bmatrix}
X_{0,\ell}(q_\perp,\varphi) \\
\Pi_{0,\ell}^*(q_\perp,\varphi)
\end{bmatrix} \, ,
\label{eq:bogo_ansatz}
\end{equation}
where $\Omega_\ell(q_\perp)$ is the quasifrequency and the amplitudes $X_{0,\ell}$ and $\Pi_{0,\ell}^*$ are themselves $\pi$-periodic in $\varphi$.
In order to determine these solutions we adopt Hill's method~\cite{Deconinck2006}. First, we formally represent $\mathcal{B}$ as a Fourier expansion,
$\mathcal{B} = \sum_{n = - \infty}^{+ \infty} \mathcal{B}_n \me^{- 2 \mi n \varphi}$, where the matrix expansion coefficients $\mathcal{B}_n = 0$
when $n \neq 0, \pm 1$, and $\mathcal{B}_{-1} = - \mathcal{B}_1^*$. Similarly, we Fourier expand the amplitudes entering the
Ansatz~\eqref{eq:bogo_ansatz}, $X_{0,\ell}(q_\perp,\varphi) = \sum_{n = - \infty}^{+ \infty} X_{0,\ell,n}(q_\perp) \me^{- 2 \mi n \varphi}$ and
$\Pi_{0,\ell}^*(q_\perp,\varphi) = \sum_{n = - \infty}^{+ \infty} \Pi_{0,\ell,n}^*(q_\perp) \me^{- 2 \mi n \varphi}$, where the
$X_{0,\ell,n}(q_\perp)$'s and $\Pi_{0,\ell,n}^*(q_\perp)$'s are the $4$-component expansion coefficients. Plugging these expansions into
Eq.~\eqref{eq:bogo_eq} and equating the terms on the two sides having the same oscillating behavior one finds
\begin{equation}
\sum_{n' = - \infty}^{+ \infty} \mathcal{F}_{n n'}
\begin{bmatrix}
X_{0,\ell,n'} \\
\Pi_{0,\ell,n'}^*
\end{bmatrix}
= \Omega_\ell
\begin{bmatrix}
X_{0,\ell,n} \\
\Pi_{0,\ell,n}^*
\end{bmatrix} \, .
\label{eq:bogo_eq_exp}
\end{equation}
Here we have introduced the infinite-dimensional matrix $\mathcal{F}$ whose entries $\mathcal{F}_{n n'} = \mathcal{B}_{n-n'} - n \Delta k \mathbb{I}_8
\delta_{n,n'}$ are themselves $8 \times 8$ matrices, and $\mathbb{I}_8$ the $8 \times 8$ identity matrix. Then, Eq.~\eqref{eq:bogo_eq_exp} is the
eigenvalue problem for $\mathcal{F}$. One can immediately see that for each solution with quasifrequency $\Omega_\ell$ there exists another one having
quasifrequency $\Omega_\ell + n \Delta k$ for arbitrary $n \in \mathbb{Z}$. However, all these infinitely many solutions of Eq.~\eqref{eq:bogo_eq_exp}
are physically equivalent as they corresponds to a single Bogoliubov mode of the form~\eqref{eq:bogo_ansatz}. For this reason, one can restrict the
real part of $\Omega_\ell$ in the first Brillouin zone, $-|\Delta k|/2 < \Re \Omega_\ell \leq |\Delta k|/2$, as done in Fig.~2 of the main text.
In addition, from the identity $\mathcal{F}_{-n, -n'} = - \mathcal{F}_{n n'}^*$ it follows that both $\Omega_\ell$ and $- \Omega_\ell^*$ belong to the
spectrum of $\mathcal{F}$. Finally, the relation
\begin{equation}
\begin{bmatrix}
0 & \mi \mathbb{I}_4 \\
- \mi \mathbb{I}_4 & 0
\end{bmatrix}
\mathcal{F}_{n n'}
\begin{bmatrix}
0 & \mi \mathbb{I}_4 \\
- \mi \mathbb{I}_4 & 0
\end{bmatrix}^{-1}
= \mathcal{F}_{-n, -n'}^\dagger \, ,
\label{eq:bogo_mat_symm}
\end{equation}
with $\mathbb{I}_4$ the $4 \times 4$ identity matrix, ensures that $\mathcal{F}$ and $\mathcal{F}^\dagger$ have the same spectrum, meaning that
quasifrequencies occur in complex conjugate pairs. Combining Eqs.~\eqref{eq:bogo_eq_exp} and~\eqref{eq:bogo_mat_symm}, after a few appropriate
manipulations, we obtain the amplitude orthonormalization relations
\begin{equation}
\sum_{n' = - \infty}^{+ \infty} \mi \left[X_{0,\ell',n'}^\dagger(q_\perp) \Pi_{0,\ell,n'+n}^*(q_\perp)
- \Pi_{0,\ell',n'}^T(q_\perp) X_{0,\ell,n'+n}(q_\perp) \right]
= \mathcal{N}_\ell(q_\perp) \delta_{\ell' \bar{\ell}} \delta_{n,0} \, ,
\label{eq:bogo_amp_norm}
\end{equation}
where $\bar{\ell}$ labels the solution with quasifrequency $\Omega_{\bar{\ell}} = \Omega_\ell^*$ and $\mathcal{N}_\ell$ is the mode norm.
From Eq.~\eqref{eq:bogo_amp_norm}, we easily obtain the relation
\begin{equation}
\mi [X_{0,\ell}^\dagger(q_\perp,\varphi) \Pi_{0,\ell'}^*(q_\perp,\varphi)
- \Pi_{0,\ell}^T(q_\perp,\varphi) X_{0,\ell'}(q_\perp,\varphi)] = \mathcal{N}_\ell(q_\perp)\delta_{\ell'\bar{\ell}} \, ,
\label{eq:bogo_amp_norm_phi}
\end{equation}
holding at arbitrary $\varphi$. Then, taking Eq.~(6) of the main text at $z=0$ and projecting it onto each separate mode by means of
Eq.~\eqref{eq:bogo_amp_norm_phi} at $z=0$, we end up with an expression for the weights as functions of the input field:
\begin{equation}
C_\ell(\vec{q}_\perp)
= \frac{\mi}{\mathcal{N}_\ell(q_\perp)}
\left[ X_{0,\bar{\ell}}^\dagger(q_\perp,\varphi_q) \Pi^*(\vec{q}_\perp,z=0)
- \Pi_{0,\bar{\ell}}^T(q_\perp,\varphi_q) X(\vec{q}_\perp,z=0) \right] \, .
\label{eq:bogo_weight}
\end{equation}

Up to now no approximation has been made. In order to perform the numerical computation of the Bogoliubov spectrum and amplitudes, the Hill method
prescribes one to choose a (sufficiently large) positive integer $N$ and truncate $\mathcal{F}_{n n'}$ to the square block with
$-(2N+1) \leq n,n' \leq 2N+1$. After checking the convergence of the numerical results,  we have chosen $N = 20$ for the figures of the Letter.

\section{Parametric resonance conditions}
\label{sec:param_reson}
The most basic model in which one can observe a parametric resonance is represented by a harmonic oscillator whose frequency is weakly modulated
in time. This system is called parametric oscillator and is described by the differential equation
\begin{equation}
\frac{\dif^2 f}{\dif t^2} + \omega_0^2 (1 + \varepsilon \cos \omega t) f = 0 \, ,
\label{eq:param_osc_1c}
\end{equation}
where $0 < \varepsilon \ll 1$, $\omega$ is the driving frequency, and $\omega_0$ is the natural frequency at which the function $f$ oscillates
in the absence of the driving. It is well known (see, e.g., Ref.~\cite{Landau_Lifshitz_01_book}) that Eq.~\eqref{eq:param_osc_1c} has oscillating
solutions that exponentially grow in time when $\omega$ takes values in an interval about $2 \omega_0 / n$, with $n$ a positive integer. The width
of the interval and the corresponding growth rate of the solutions decreases with $n$. Parametric resonances appear even if one adds a friction term
$- 2 \lambda \frac{\dif f}{\dif t}$ to the oscillator equation~\eqref{eq:param_osc_1c}, provided that $\varepsilon$ exceeds a critical value
$\varepsilon_\mathrm{cr}$ that grows with $n$.

In our fluid of light, the time evolution of the perturbation field can be mapped to a multicomponent parametric oscillator, where the drive originates
from the nonparaxial terms of the Helmholtz equation, rather than being externally imposed. This mapping is easily accomplished by looking at the
Euler-Lagrange equations $\frac{\dif}{\dif z} \frac{\partial \tilde{\mathcal{L}}^{(2)}}{\partial \dot{X}^\dagger}
- \frac{\partial \tilde{\mathcal{L}}^{(2)}}{\partial X^\dagger} = 0$ for the quadratic Lagrangian $\tilde{\mathcal{L}}^{(2)}$ [see Eq. (4) of the main
text], which read
\begin{equation}
\ddot{X} + \Lambda_2^{-1} (\dot{\Lambda}_2 + \Lambda_1 - \Lambda_1^T) \dot{X} + \Lambda_2^{-1} (\dot{\Lambda}_1 + \Lambda_0) X = 0 \, .
\label{eq:param_osci_mc}
\end{equation}
Notice the presence of a time-oscillating friction term proportional to $\dot{X}$. Equation~\eqref{eq:param_osci_mc} can be rewritten in the
form~\eqref{eq:bogo_eq}--\eqref{eq:bogo_mat} after switching to the Hamiltonian formalism and introducing the conjugate momenta $\Pi$. In this framework,
the conditions for parametric resonances involves the eigenvalues of the time average $\mathcal{B}_0 = \frac{\Delta k}{2 \pi}
\int_{- \pi / \Delta k}^{\pi / \Delta k} \dif z \, \mathcal{B}(z)$ of the Bogoliubov matrix $\mathcal{B}$ \cite{Lellouch2017}. At moderate $q_\perp$ there
are four positive eigenvalues, denoted by $\Omega_a^{(0)}(q_\perp)$ with $a = 1, 2, 3, 4$ (the negative eigenvalues being just their opposites), which
play the role of the natural oscillation frequencies of our system. Then, parametric resonances are expected to occur at $q_\perp$-values such that the
two lowest frequencies $\Omega_{1,2}^{(0)}(q_\perp)$ fulfill conditions of the kind $\Delta k \simeq 2 \Omega_1^{(0)}(q_\perp) / n$, $\Delta k \simeq
2 \Omega_2^{(0)}(q_\perp) / n$, or $\Delta k \simeq [\Omega_1^{(0)}(q_\perp) + \Omega_2^{(0)}(q_\perp)] / n$, with $n$ a positive integer [the highest
frequencies $\Omega_3^{(0)}(q_\perp)$ and $\Omega_4^{(0)}(q_\perp)$ are always much larger than $\Delta k$ and thus are not associated with any resonance].
We have checked that each peak of finite imaginary part in the spectrum of Fig.~2(b) of the main text indeed emerges in correspondence with one of these
conditions, see labels on the figure. In addition, comparing Fig.~2(a) and Fig.~2(b) of the main text one sees that each parametric resonance is also
associated with the crossing of two branches of the real part of the quasifrequency spectrum; hence, these points can be regarded as avoided crossings
in the complex quasifrequency plane. Finally, we point out that, because of the friction term in Eq.~\eqref{eq:param_osci_mc}, only a finite number of
resonances characterized by small values of the integer index $n$ can effectively occur.

\section{Input conjugate momenta}
\label{sec:input_conj_mom}
The study of a nonparaxial light beam propagating inside a bulk nonlinear medium requires one to solve the Bogoliubov equations~\eqref{eq:bogo_eq} for
a given input field profile $X(\vec{q}_\perp,z=0)$, such as that of Eq.~(7) of the main text. However, this is not enough to uniquely determine the
solution to the problem, as the input value of the conjugate momenta $\Pi^*(\vec{q}_\perp,z=0)$ is also required. In order to fix this value, we first
recall that the Bogoliubov spectrum of the system includes eight branches, only half of which are associated with field modes propagating forward
inside the medium; the remaining branches describe modes propagating backward in effective time, i.e., along the negative $z$ direction, which
do not appear in the paraxial framework. To know whether a given branch $\ell$ propagates forward or backward we define an average frequency
$\bar{\Omega}_\ell = \Omega_\ell + \mathcal{N}_\ell^{-1} \Delta k \sum_{n = - \infty}^{+ \infty} \mi n \left(X_{0,\bar{\ell},n}^\dagger \Pi_{0,\ell,n}^*
- \Pi_{0,\bar{\ell},n}^T X_{0,\ell,n} \right)$, which can be regarded as the expectation value of the effective energy operator $\mi\nabla_z$ [cfr. the
orthonormalization condition~\eqref{eq:bogo_amp_norm}]. Forward (backward) modes are those having the smallest (largest) value of
$|\Re \bar{\Omega}_\ell|$. We take $\Pi^*(\vec{q}_\perp,z=0)$ such that only the forward modes, whose frequencies will be denoted by $\Omega_{-,d}$
and $\Omega_{-,s}$ (where $d$ and $s$ stand for ``density'' and ``spin'', respectively), be excited; conversely, their backward counterparts of
quasifrequency $\Omega_{+,d}$ and $\Omega_{+,s}$ do not appear in the solution of Eq.~\eqref{eq:bogo_eq}.

The practical implementation of the above criterion requires as a preliminary step to consider the operator evolving the solution of the Bogoliubov
equations~\eqref{eq:bogo_eq} with respect to effective time by one oscillation period $2\pi / \Delta k$,
\begin{equation}
\begin{bmatrix}
X(\vec{q}_\perp,z=2\pi/\Delta k) \\
\Pi^*(\vec{q}_\perp,z=2\pi/\Delta k)
\end{bmatrix}
=
\mathcal{U}(2\pi/\Delta k,0)
\begin{bmatrix}
X(\vec{q}_\perp,z=0) \\
\Pi^*(\vec{q}_\perp,z=0)
\end{bmatrix} \, .
\label{eq:inp_evol_XP}
\end{equation}
One can decompose this operator as $\mathcal{U}(2\pi/\Delta k,0) = \mathcal{P} \mathcal{U}_D \mathcal{P}^{-1}$, with
\begin{equation}
\mathcal{U}_D =
\begin{bmatrix}
\me^{- 2 \pi \mi \mathcal{B}_{D+} / \Delta k} & 0 \\
0 & \me^{- 2 \pi \mi \mathcal{B}_{D-} / \Delta k}
\end{bmatrix} \, ,
\qquad
\mathcal{P} =
\begin{bmatrix}
\mathcal{P}_{X+} & \mathcal{P}_{X-} \\
\mathcal{P}_{\Pi+} & \mathcal{P}_{\Pi-}
\end{bmatrix} \, ,
\qquad
\mathcal{P}^{-1} =
\begin{bmatrix}
\left(\mathcal{P}^{-1}\right)_{X+} & \left(\mathcal{P}^{-1}\right)_{\Pi+} \\
\left(\mathcal{P}^{-1}\right)_{X-} & \left(\mathcal{P}^{-1}\right)_{\Pi-}
\end{bmatrix} \, .
\label{eq:inp_diag_mat}
\end{equation}
The $4 \times 4$ blocks entering $\mathcal{U}_D$ and $\mathcal{P}$ are given by
\begin{subequations}
\label{eq:inp_mat_block}
\begin{align}
\mathcal{B}_{D\pm} &{}
= \operatorname{diag}\left( \Omega_{d,\pm}(q_\perp), -\Omega_{d,\pm}^*(q_\perp), \Omega_{s,\pm}(q_\perp), -\Omega_{s,\pm}^*(q_\perp) \right) \, ,
\label{eq:inp_mat_block_bogo} \\
\mathcal{P}_{X\pm} &{} =
\begin{bmatrix}
X_{d,\pm}(\vec{q}_\perp,z=0) & X_{d,\pm}^*(-\vec{q}_\perp,z=0) & X_{s,\pm}(\vec{q}_\perp,z=0) & X_{s,\pm}^*(-\vec{q}_\perp,z=0)
\end{bmatrix} \, ,
\label{eq:inp_mat_block_X} \\
\mathcal{P}_{\Pi\pm} &{} =
\begin{bmatrix}
\Pi_{d,\pm}^*(\vec{q}_\perp,z=0) & \Pi_{d,\pm}(-\vec{q}_\perp,z=0) & \Pi_{s,\pm}^*(\vec{q}_\perp,z=0) & \Pi_{s,\pm}(-\vec{q}_\perp,z=0)
\end{bmatrix} \, .
\label{eq:inp_mat_block_Pi}
\end{align}
\end{subequations}
Notice that the columns of $\mathcal{P}$ correspond to the eight wave functions defined in Eq.~\eqref{eq:bogo_ansatz} taken at $z=0$.
Consequently, the blocks of $\mathcal{P}^{-1}$ can be found from the orthonormalization conditions discussed in Sec.~\ref{sec:bogo_hill}. We now
define the new variables
\begin{equation}
\begin{bmatrix}
Y_{D+} \\
Y_{D-}
\end{bmatrix}
= \mathcal{P}^{-1}
\begin{bmatrix}
X \\
\Pi
\end{bmatrix} \, ,
\label{eq:inp_def_Y}
\end{equation}
whose effective-time evolution over one oscillation period is simply given by a phase factor, $Y_{D\pm}(\vec{q}_\perp,z=2\pi / \Delta k)
= \me^{- 2 \pi \mi \mathcal{B}_{D\pm} / \Delta k} Y_{D\pm}(\vec{q}_\perp,z=0)$. Requiring $Y_{D+}(\vec{q}_\perp,z=0) = 0$ is a sufficient condition to
ensure that the backward Bogoliubov modes are not excited. When expressing this constraint in terms of the original variables one obtains
\begin{equation}
\Pi^*(\vec{q}_\perp,z=0) = - \left[ \left(\mathcal{P}^{-1}\right)_{\Pi+} \right]^{-1} \left(\mathcal{P}^{-1}\right)_{X+} X(\vec{q}_\perp,z=0) \, .
\label{eq:inp_init_cond_Pi}
\end{equation}
We checked that in the paraxial limit this condition is consistent with the relation~\eqref{eq:paraxial_el_eq}, connecting $X$ to its first-order
derivative, taken at $z = 0$.

\section{Intensity-intensity correlation function}
\label{sec:g2}
In order to compute the intensity-intensity correlation function one has to start from the input field [Eq.~(7) of the main text] and write down
the corresponding expression of the four-component vector $X$,
\begin{equation}
X(\vec{q}_\perp,z=0) = \epsilon \sum_{\alpha = r,i} \sum_{\sigma = \pm} X_{\alpha,\sigma} \tilde{\phi}_{\alpha,\sigma}(\vec{q}_\perp) \, ,
\label{eq:speckle_in_X}
\end{equation}
where
\begin{equation}
X_{r,+} =
\begin{bmatrix}
\cos\frac{\vartheta_0}{2} \\
-\sin\frac{\vartheta_0}{2} \\
0 \\
0
\end{bmatrix}  \, ,
\quad
X_{r,-} =
\begin{bmatrix}
\sin\frac{\vartheta_0}{2} \\
\cos\frac{\vartheta_0}{2} \\
0 \\
0
\end{bmatrix}  \, ,
\quad
X_{i,+} =
\begin{bmatrix}
0 \\
0 \\
\frac{1}{2\cos\frac{\vartheta_0}{2}} \\
\frac{1}{2\cos\frac{\vartheta_0}{2}}
\end{bmatrix}  \, ,
\quad
X_{i,-} =
\begin{bmatrix}
0 \\
0 \\
\frac{1}{2\sin\frac{\vartheta_0}{2}} \\
-\frac{1}{2\sin\frac{\vartheta_0}{2}}
\end{bmatrix}  \, .
\label{eq:speckle_in_X_as}
\end{equation}
The correlators of the Fourier transform of the speckle fields are given by
$\langle \tilde{\phi}_{r,\sigma}(\vec{q}_\perp) \tilde{\phi}_{r,\sigma'}^*(\vec{q}_\perp')\rangle
= \langle \tilde{\phi}_{i,\sigma}(\vec{q}_\perp) \tilde{\phi}_{i,\sigma'}^*(\vec{q}_\perp')\rangle
= (2\pi)^2 \delta(\vec{q}_\perp' - \vec{q}_\perp) \tilde{\gamma}(q_\perp) \delta_{\sigma\sigma'} / 2$
and $\langle \tilde{\phi}_{r,\sigma}(\vec{q}_\perp) \tilde{\phi}_{i,\sigma'}^*(\vec{q}_\perp')\rangle = 0$, with $\tilde{\gamma}(q_\perp) =
4 \pi \sigma^2 \exp(- \sigma^2 q_\perp^2)$. From the discussion of Sec.~\ref{sec:input_conj_mom}, we know that the input conjugate momenta take the
form
\begin{equation}
\Pi^*(\vec{q}_\perp,z=0) =
\epsilon \sum_{\alpha = r,i} \sum_{\sigma = \pm} \Pi_{\alpha,\sigma}^*(\vec{q}_\perp) \tilde{\phi}_{\alpha,\sigma}(\vec{q}_\perp) \, ,
\label{eq:speckle_in_Pi}
\end{equation}
where the $\Pi_{\alpha,\sigma}^*$'s can be directly deduced from Eq.~\eqref{eq:inp_init_cond_Pi}. After solving the problem as discussed in the
previous sections and in the main text, one can write the intensity fluctuation at arbitrary $\vec{r}_\perp$ and $z$ as
\begin{equation}
\delta I(\vec{r}_\perp,z) = \int \frac{\dif^2 q_\perp}{(2\pi)^2}
\sum_\ell C_\ell(\vec{q}_\perp) \delta I_{0,\ell}(q_\perp,\varphi(z))
\me^{\mi[\vec{q}_\perp \cdot \vec{r}_\perp - \Omega_\ell(q_\perp) z]} \, ,
\label{eq:speckle_sol_int}
\end{equation}
where $\delta I_{0,\ell}$ is (up to a factor $2 I_0$) the first component of the amplitude vector $X_{0,\ell}$ [see Eq.~\eqref{eq:bogo_ansatz}] and
the weights are given by Eq.~\eqref{eq:bogo_weight}. To evaluate $g_2(z)\equiv \langle \delta I(\vec{r}_\perp,0) \delta I(\vec{r}_\perp,z) \rangle$
one needs the expression of the weight correlator
\begin{equation}
\langle C_\ell(\vec{q}_\perp) C_{\ell'}^*(\vec{q}_\perp') \rangle
= \epsilon^2 (2\pi)^2 \delta(\vec{q}_\perp' - \vec{q}_\perp) \tilde{\gamma}(q_\perp) \frac{\Xi_{\ell\ell'}^*(\vec{q}_\perp)}{2} \, ,
\label{eq:speckle_weight_corr}
\end{equation}
where
\begin{equation}
\begin{split}
\Xi_{\ell\ell'}^*(\vec{q}_\perp)
= {}&{} \sum_{\alpha = r,i} \sum_{\sigma = \pm}
\left[ X_{0,\bar{\ell}}^\dagger(q_\perp,\varphi_q) \Pi_{\alpha,\sigma}^*(\vec{q}_\perp)
- \Pi_{0,\bar{\ell}}^T(q_\perp,\varphi_q) X_{\alpha,\sigma} \right] \\
&{} \times \left[ X_{0,\bar{\ell}'}^\dagger(q_\perp,\varphi_q) \Pi_{\alpha,\sigma}^*(\vec{q}_\perp)
- \Pi_{0,\bar{\ell}'}^T(q_\perp,\varphi_q) X_{\alpha,\sigma} \right]^*
\left[ \mathcal{N}_\ell(q_\perp) \mathcal{N}_{\ell'}^*(q_\perp) \right]^{-1} \, .
\end{split}
\label{eq:speckle_weight_corr_coeff}
\end{equation}
Then, a straightforward calculation yields
\begin{equation}
\begin{split}
g_2(z) = \epsilon^2 I_0^2 \sum_\ell \int_0^{\infty} \frac{\dif q_\perp}{2\pi} \, q_\perp \tilde{\gamma}(q_\perp) K_\ell(q_\perp, z)
\me^{- \mi \Omega_\ell(q_\perp) z} \, ,
\end{split}
\label{eq:speckle_2corr}
\end{equation}
where we have introduced the modulated coefficients
\begin{equation}
K_{\ell}(q_\perp,z) =
\frac{1}{2 I_0^2} \int_{- \pi}^\pi \frac{\dif\varphi_q}{2\pi}
\sum_{\ell'} \Xi_{\ell'\ell}(\vec{q}_\perp)
\delta I_{0,\ell'}^*(q_\perp,\varphi_q) \delta I_{0,\ell}(q_\perp,\varphi(z)) \, .
\label{eq:speckle_2corr_coeff}
\end{equation}
Equation~\eqref{eq:speckle_2corr} also holds in the paraxial framework, with the coefficients taking the $q_\perp$- and $z$-independent values
\begin{equation}
K_{d(s)} = \frac{1}{2} \pm \frac{g_d + g_s \cos 2 \vartheta_0}{2\sqrt{g_d^2 + g_s^2 + 2 g_d g_s \cos 2\vartheta_0}} \, ,
\label{eq:speckle_2corr_coeff_par}
\end{equation}
where the upper (lower) sign refers to the two density (spin) branches of the Bogoliubov spectrum. In addition, at large $z$ one can approximate
$\Omega_{d(s)}(q_\perp) \simeq c_{d(s)} q_\perp$ in the integral~\eqref{eq:speckle_2corr}, with $c_{d(s)}$ the paraxial sound
velocities~\eqref{eq:paraxial_sound}, yielding the asymptotic behavior $g_2(z) \simeq - \epsilon^2 I_0^2 (\sqrt{g_d I_0/\beta_0} {z}/{2 \sigma})^{-2}$.

\begin{figure}
\includegraphics[scale=1]{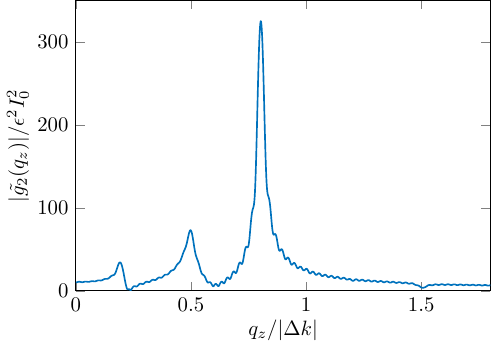}
\caption{Discrete Fourier transform of $g_2$ (in modulus) as a function
of the effective frequency $q_z$ (we can restrict to $q_z > 0$ because
of symmetry). The parameters are the same as the blue solid curves of
Figs.~3 and 4(a) of the main text, namely the background polarization
$\vartheta_0 = \pi / 4$, the nonlinear couplings $g_d I_0 / \beta_0 = 0.2$,
$g_s I_0 / \beta_0 = 0.05$, and the correlation length $\beta_0\sigma = 15$.}
\label{fig:g2_fourier}
\end{figure}

In the nonparaxial regime the $K_\ell$'s are periodic in $z$, giving rise to the peculiar oscillating behavior shown in Figs.~3 and~4(a) of the main
text. To identify
the dominant oscillation frequencies we numerically evaluate $g_2(z)$ in a window of fairly large $z$ and then compute its
(discrete) Fourier transform $\tilde{g}_2(q_z)$.
At large $\sigma$ we find two symmetric peaks at some $q_z$ close to $\pm |\Delta k|$, meaning that
the oscillations are practically harmonic. However, as $\sigma$ is
decreased (at fixed values of the other parameters) additional peaks occur, and the
oscillations become more and more anharmonic. As an example, in
Fig.~\ref{fig:g2_fourier} we show the Fourier transform of $g_2$ evaluated in the
$1000 \, z_\mathrm{NL} \leq z \leq 2000 \, z_\mathrm{NL}$ range with the same
parameters as the blue curves of Figs.~3 and~4(a) of the main text. One
can clearly see the main peaks at $q_z \simeq \pm 0.8 \, |\Delta k|$ and the secondary peaks at
$q_z \simeq \pm 0.5 \, |\Delta k|$ and
$q_z \simeq \pm 0.2 \, |\Delta k|$. However, we noticed that, each time a new peak appears, when further lowering $\sigma$ its position
remains
basically $\sigma$-independent.